\documentstyle[11pt,aaspp4,flushrt,tighten]{article}
\input epsf.sty
{\vskip1truecm}
\def\mathnew{\mathsurround=0pt}
\font\trm=cmr10

\def\simov#1#2{\lower .5pt\vbox{\baselineskip0pt \lineskip-.5pt
       \ialign{$\mathnew#1\hfil##\hfil$\crcr#2\crcr\sim\crcr}}}
\def\gtsima{$\; \buildrel > \over \sim \;$}
\def\simgreat{\mathrel{\mathpalette\simov >}}
\def\simless{\mathrel{\mathpalette\simov <}}

\def\bd{{\bf d}}

\def\bm{{\bf m}}
\def\Pr{{\cal P}}
\def\cf{{\it cf.\ }}
\def\eg{{\it e.g.\ }}
\def\ie{{\it i.e.\ }}
\def\etal{{\it et al.\ }}
\def\hmpc{\ {\rm h^{-1}Mpc}}
\def\kms{\ {\rm km\,s^{-1}}}
\def\RM{{\rm RM}}
\def\degm2{\, {\rm deg\,m}^{-2}}
\def\radm2{\, {\rm rad\,m}^{-2}}

\def\rmgs{\widehat{\RM}_G({\hat {z}})}

\def\today{\ifcase\month\or
  January\or February\or March\or April\or May\or June\or
  July\or August\or September\or October\or November\or
December\fi
  \space\number\day, \number\year}

\begin{document}


\title{Determination of the Primordial Magnetic Field  Power Spectrum \\
by Faraday Rotation Correlations}
\author{Tsafrir Kolatt$^1$}
\affil{Harvard-Smithsonian Center for Astrophysics, 60 Garden St.,
Cambridge, MA 02138}
\affil{and}
\affil{The Physics Department and Lick Observatory, UCSC, Santa-Cruz, CA
95064}
\altaffiltext{1}{email:tsafrir@physics.ucsc.edu}

\begin{abstract} 
This paper introduces the formalism which connects between rotation
measure ($\RM$) measurements for extragalactic sources and the cosmological
magnetic field power spectrum. It is shown that
the amplitude and shape of the cosmological magnetic field power
spectrum can be constrained by using a few hundred radio
sources, for which Faraday $\RM$s are available.
This constraint is of the form 
$B_{rms} \simless 1 \times [2.6\times10^{-7} {\rm cm}^{-3}/ \bar n_b] h $ 
nano-Gauss (nG) 
on $\sim 10-50 \hmpc$  scales, with $\bar n_b$ the average baryon
density and $h$ the Hubble parameter in $100$ km s$^{-1}$ Mpc$^{-1}$
units. The constraint is superior to and supersedes
any other constraint which come from either CMB fluctuations, Baryonic nucleosynthesis, 
or the first two multipoles of the magnetic field expansion. 
The most adequate method for the constraint calculation uses the
Bayesian approach to the maximum likelihood function.
I demonstrate the
ability to detect such magnetic fields by constructing 
simulations of the field and mimicking observations. This procedure also
provides error estimates for the derived quantities. 

The two main noise contributions due to the Galactic RM and the internal
RM are treated in a statistical way following an evaluation of their
distribution.
For a range of magnetic field power spectra
with power indices $-1\leq n \leq 1$
in a flat cosmology ($\Omega_m$=1) 
we estimate the signal-to-noise ratio, $Q$, for limits on the magnetic
field $B_{rms}$ on $\sim 50\hmpc$ scale. Employing one patch of a few
square degrees on the sky with source number density $n_{src}$, an
approximate estimate yields
$Q\simeq 3 \times (B_{rms}/1 {\rm nG})(n_{src}/50\, {\rm deg}^{-2})
(2.6\times10^{-7} {\rm cm}^{-3}/ \bar n_b)\, h $. 
An all sky coverage, with much sparser, but carefully tailored sample of 
$\sim 500$ sources, yields $Q\simeq 1$ with the same scaling. An ideal
combination of small densely sampled patches and sparse all-sky coverage
yields $Q\simeq 3$ with better constraints for the power index.
All of these estimates are corroborated by the simulations.
\end{abstract}

\section{INTRODUCTION}
\label{sec:intro}
There is plenty evidence for magnetic fields on the scale of the earth, the
solar system, the interstellar medium, galaxies and clusters of galaxies [for
the extragalactic magnetic field see review by Kronberg (1994, K94) and references
therein].  
We still don't know however whether any larger scale magnetic fields 
exist. There are various upper limits on the rms value of the
cosmological magnetic field, but the upper limits are still too high
to exclude any relevance to magnetic fields observed 
on smaller scales.
Cosmological magnetic fields might help to resolve the bothersome question
of the origin of the $\sim 10^{-6}$G magnetic field which exists on galactic scales
or fields of similar size seen on the scales of galaxy clusters. A primordial
(pre -- galaxy formation) magnetic
field amplified by the gravitational collapse, and then by processes like
differential rotation and dynamo amplification (Parker 1979), can serve to seed
these observed smaller scale magnetic fields.

The most stringent upper limits to date for magnetic fields on cosmological 
scales come from three different sources:
\begin{enumerate}
\item  Big Bang Baryonic Nucleosynthesis (BBN).

Magnetic fields that existed during the BBN epoch would affect the expansion
rate, the reaction rates, the electron density (and possibly the
space-time geometry). By taking all these
effects into account in the calculation of the element abundances,
and then comparing the results with the observed abundances
one can set limits on the magnetic field amplitude.
The limits for homogeneous magnetic fields on scales $\gg 10^{-14}
\hmpc$ (distances are quoted in comoving coordinates)
and up to the BBN horizon size ($\sim 10^{-4} \hmpc$)
are $10^{-9} - 10^{-6}$ G in terms of today's values (Grasso \& Rubinstein 1995, 1996
; Cheng, Olinto, Schramm, \&  Truran 1996).
The relevance of these limits for the maximum value of magnetic field seeds
on the subgalactic scale is obvious, but in order for these limits to be
relevant to the intergalactic magnetic field,
further assumptions about the
super-horizon magnetic field power spectrum and the magnetic field
generation epoch should be made.

\item  The CMB radiation.

The presence of magnetic fields during the time of decoupling causes
anisotropy, \ie different expansion rates in different directions
(Zel'dovich \& Novikov 1975). Masden (1989) provides a useful formula
for the connection between the measured temperature fluctuations on a
given scale at $z_d$, the decoupling time, and the limits on the
equivalent scale today. For a magnetic field energy density much
smaller than the matter energy density, the limit is
\begin{equation}
\label{eq:cmb_lim}
B_{rms}(R) \le 3\times 10^{-4} h {\sqrt {{\Delta T \over T  } (R) \Omega} \over \sqrt {1+z_d} }
\,\,\, {\rm  G} \,. 
\end{equation}
For typical values of $h=1$ (the Hubble constant in units of $100$ km
s$^{-1}$ Mpc$^{-1}$), temperature fluctuations of $\sim 10^{-5}$, $\Omega =
1$ and $z_d \simeq 10^{3}$ the limit becomes $B_{rms} < 10^{-8} - 10^{-9}$G.
Recently, Barrow, Frreira, \& Silk (1997) used the COBE 4-year data to
constrain the magnetic field on the horizon scale by $B(R=R_h) < 6.8
(\Omega h^2)^{1/2}$ nG.

\item Multipole expansion of $\RM$ observations.

Passing through a magnetoionic medium, polarized light undergoes
rotation of the polarization vector (Faraday rotation, for definition see
\S\ref{sec:corr}). The rotation amount is proportional to the dot product of
the magnetic field and the light propagation direction, and to the 
distance the light travels through the medium. 
If any cosmological magnetic field
exists, the farther away the source, the more the polarized light
component is rotated. We refer to this as the ``monopole" term.
This term exists even if the field has no preferred direction on
the horizon scale, namely even without breaking the isotropy hypothesis.

If there is a preferred direction to the field on a cosmological scale,
a dipole will show up in the $\RM$ measurements across the sky,
due to the different sign of the dot product (and thus the
different rotation direction).
So far attempts have been made only to identify a monopole or a dipole in the magnetic
field for scales of $z\simeq2.5$ (Kronberg 1976) and $z\simeq 3.6$
(Vall\'ee 1990). 
The search for a dipole signature in the
Faraday rotation values for a sample of extra-galactic sources (QSO)
yields limits of $B < 1 - 10^{-1}$ nG depending on the assumed
cosmology (Kronberg \& Simard-Normandin 1976, Kronberg 1976).
Vall\'ee (1990) tried to estimate the limits on the dipole and the
monopole terms and concluded that an all-prevailing (up to $z=3.6$) 
field must be smaller
than $6\times10^{-2} \left[10^{-6} {\rm  cm}^{-3} \over  \bar n_b\right]$ nG .

\end{enumerate}

\noindent Other potential methods of detecting a possible cosmological magnetic
field include
distortion of the acoustic (``Doppler") peak in the CMB power spectrum
(Adams, Danielsson, Grasso, \& Rubinstein 1996)
and Faraday rotation of the CMB radiation polarized components 
(Loeb \& Kosowsky 1996). These methods where not designed to probe any
magnetic field generated after recombination.
Their implementation is still pending on the upcoming measurements of the CMB fluctuations and polarization.

A method that does probe magnetic fields generated in the post
recombination era, relies on cosmic ray (CR) detection.
The effect of magnetic fields on high energy CRs
($>10^{18}$ eV) is twofold. It will first alter the energy distribution
of CRs (Lee, Olinto, \& Sigl 1995, Waxman \& Miralda-Escud\'e 1996)
and then, if the CR direction is identified and attributed to a known
region or source, the magnetic field either smears the directionality
(for field coherent scales much smaller than the distance to the source) or
deflects the CR direction from aligning with the source direction. 
Current limits from this last effect are either very weak or non-existing.

Theories for primordial magnetic fields in the framework of structure
evolution paradigms have been suggested by a number of authors. 
Magnetic fields on cosmological scales emerge in the framework of
these theories in one of three epochs: the inflation era, the plasma
era, and the post-recombination era. In the inflation era,
magnetic fields form due to quantum mechanical processes (Turner \&
Widrow 1988; Quashnock, Loeb,
\& Spergel 1989, Vachaspati 1991, Ratra 1992a, 1992b, Dolgov \& Silk 1993) or
possibly magneto-hydrodynamics (Brandenburg, Enquist, \& Olesen 1996). During
the plasma era, prior to recombination, magnetic fields form due to
vorticity caused by the mass difference between the electron and the
proton (Harrison 1973, but see Rees 1987), or where magnetic field
fluctuation survival (no vorticity assumed) depends on the fluctuation scale 
(Tajima \etal 1992). Other authors have pointed out that even in the post
recombination era (but before galaxy formation) magnetic fields can be
generated by tidal torques (Zweibel 1988), or by cosmic strings wakes
(Ostriker \& Thompson 1987, Thompson 1990). Magnetic field generated
in the process of structure formation itself, due to falling
matter (Pudritz \& silk 1989), or star burst regions, are less
likely to feed back into the intergalactic space and to affect the
cosmological magnetic field. 
Nevertheless, proposals in this spirit were
also considered by taking into account wind driven plasma which fills
intergalactic space with magnetic fields.

Whatever the origin of cosmological magnetic fields, each
suggested theory, specifies the magnetic field amplitude
and power spectrum shape. Thus, each of the theories can in
principle become refutable if we could measure, or put limits on, the
normalized power spectrum of the cosmological magnetic field. 

This challenge and the realization that with existing or upcoming
data, it will be possible to measure the magnetic field power spectrum
has led us to develop a method for doing so.
We argue that much better estimates (or upper limits) for the
cosmological magnetic field can be calculated
by using the RM correlation matrix.

All previous attempts to limit the cosmological magnetic
field dipole on very large scales ($z\simeq2.5 -3.6$), 
suffer from two drawbacks. From a theoretical
point of view it is difficult to reconcile the necessary anisotropy such
a dipole imposes with
other evidence which suggest global isotropy. From a practical point of view,
this test is confined to one scale, and doesn't allow measurements over
a whole range. Only by extending the measurement to a whole range, can
one estimate the full power spectrum (PS).
Although contribution from a ``random walk"
can be useful even if only a monopole is calculated (see below),
the information is partial, inferior to the full derivation of 
the magnetic field by $\RM$ correlations, and sensitive to evolutionary effects.

Why is the correlation approach more advantageous than other type of
statistics and
in particular why is it better than the monopole approach? 
In the monopole approach we are
looking for a correlation between RM values and the source redshift.
Let's pretend at first that all measurements are exact and there are no
noise sources (we'll notice in \S\ref{sec:calc} that these two assumptions are far
too optimistic). Consider a scale $l_0$ over which the magnetic field is
coherent with an rms value of $B_{rms}$. 
A line-of-sight to a source at a distance $r_{src}$ will cross
$N_l=r_{src}/l_0$ such regions. There is no correlation in the magnetic
field orientation between the regions, and on average each one of them
contributes $\RM^i_{rms} \propto B_{rms}l_0$.
The overall contribution due to  $N_l$ random walk steps
sums up to $\RM_{rms} = \sqrt {N_l} \RM^i_{rms} \propto
B_{rms}\sqrt {l_0r_{src}}\,$. It is thus clear
that the larger $r_{src}$, the bigger $\RM_{rms}$ we expect for an
ensemble of sources located at $r_{src}\,$.
Now let's turn to the correlation approach. Two lines-of-sight
to two adjacent sources 
(separation angle $\gamma$ and assume the same redshift for the two) are
observed for correlation between the $\RM$ values of the polarized
light
emanating from them. If $\gamma r_{src} \le l_0$ the two light rays
undergo {\it the same} rotation in each patch of length $l_0$ (for
small $\gamma$ we
neglect the difference in the $\cos(\theta)$ term between the magnetic
field and the line-of-sight). The ensemble average of the correlation term
would then be $<\RM_1\RM_2>^{1/2} =  N_l \RM^i_{rms} \propto B_{rms} r_{src}$,
with the same proportionality constant as before. In the limit of $\l_o
\rightarrow r_{src}$, the two expressions coincide, but for any value of
$l_0<r_{src}$, the signal from the correlation calculation is amplified
by the factor $\sqrt {r_{src}/l_0}$. For example, if $r_{src}= 1700
\hmpc $ $(z\simeq1)$ and $l_0=10 \hmpc$ the signal is amplified by more than
an order of magnitude. Moreover, the correlation provides information about the
power spectrum (different $\gamma$ and $r_{src}$ values) of the magnetic field that otherwise we wouldn't be able
to obtain. 
The correlation also makes it easyer to separate the noise from the
signal, unlike the monopole approach.
We will return to this approximation later on, under more
realistic considerations (\S \ref{subsec:ston}),
when we attempt to evaluate the number of pairs
needed in order to establish a certain signal-to-noise ratio at a given
scale ($\gamma$).
For now, this simple model demonstrates the advantages of this paper's
approach.

In order to
relate actual data of RM measurements to the magnetic field, 
we begin by considering 
the connection between the two.
In section \ref{sec:corr} we introduce the magnetic field correlation
tensor and its connection to the RM correlation. In section \ref{sec:calc} we
make use of these definitions and describe the calculation
procedure for the $\RM$ correlation from the raw data to the constraints
on the magnetic field power spectrum. A key issue is the estimate of noise
from non-cosmological contributions to the $\RM$.
The impatient reader can turn immediately to \S \ref{subsec:ston} to get
a rough estimate for the expected signal-to-noise ratio.
We demonstrate the procedure in
the following section (\S4), where we simulate a few realistic examples
of cosmological magnetic fields, and exploit them to get RM measurements
from which we derive back an estimate for the original power spectrum. 
The simulations also allow us to perform a realistic error analysis.
In section 5 we discuss prospects 
for applying the procedure to real data and conclude with our results.

\section{THE ROTATION MEASURE CORRELATION MATRIX}
\label{sec:corr}
Linearly polarized electro-magnetic radiation of frequency $\nu$ traveling a
distance ${\rm d}l$
through a non-relativistic plasma medium with the magnetic field ${\vec
B}$, is rotated by the angle ${\rm d} \phi$. In other
words, the polarization angle is changed by (\eg Lang 1978 Eqs. 1-268 $-$
1-270)
\begin{equation}
\label{eq:rot_ang}
{\rm d} \phi = {e^3 n_e \over 2 \pi m_e^2 \nu^2 } {\vec B} \cdot {\vec
{{\rm d} l}} \, , 
\end{equation}  
where $e$, $m_e$, and $n_e$ are the electron charge, mass, and number
density respectively (we use units of $c=1$).
The rotated polarization vector itself is written 
\begin{equation}
\label{eq:p_vec}
p(\nu) = \vert p(\nu) \vert e^{2i\phi}\, , 
\end{equation}
a pseudo-vector degenerated in rotation of $\phi$ by $n\pi$.
It is related to the Stokes parameters via
\begin{equation}
\label{eq:stokes}
\vert p \vert = {(Q^2+U^2)^{1/2} \over I}\,;\qquad \phi = {1 \over 2}
\tan^{-1}\left({U \over Q}\right)\,,
\end{equation}
that in turn are expressed by the electric field components
perpendicular to the wave propagation direction $(z)$
\[I = E_x^2 + E_y^2 \]
\[Q = E_x^2 - E_y^2 \]
\begin{equation}
\label{eq:electric}
U = E_y^*E_x + E_yE_x^*
\end{equation}
where time average is explicitly assumed.

\noindent In the cosmological context, some of the quantities are functions of
time (or equivalently redshift or distance). We assume fully ionized gas
(ionization fraction $X_e=1$) and thus the average number density of
free electrons is
\begin{equation}
\label{eq:ne}
\bar n_e=X_e \bar n_b = \bar n_b \,.
\end{equation}
The baryon number density at a specific location is $n_b(\vec x)=\bar n_b
(1+\delta(\vec x))$, with $\delta(\vec x)$ the dimensionless density
fluctuation.
Number densities scale like $a^{-3}$, where $a$ is the scale factor.
The frequency scales like $a^{-1}$, and under
the assumption of flux conservation, $B$ scales like $a^{-2}$.
The distance unit ``${\rm d}l$" in a Robertson-Walker metric is given by
\begin{equation}
\label{eq:dl_cos}
{\rm d} l = {  {\rm d} t } = { a(t) {\rm d} \eta } = {  {\rm
d} r \over  {\sqrt {1-Kr^2}} } \, ,
\end{equation}
expressed via $\eta$ the conformal time. Since $a = (1+z)^{-1}$
for the redshift $z$, the overall rotation of the polarized radiation
of a source at the redshift $z$, the direction ${\hat q}$ as observed from
${\vec x}$, (we set $x=0$ for simplicity) and at frequency $\nu_0$, is given by
\begin{equation}
\label{eq:rm_1}
\int_0^r {\rm d} \phi = {e ^3 \bar n_{b_0} \over 2
\pi m_e^2 \nu_0^2 } \int_0^{r(z)} \left[1+\delta({\hat q}r')\right]
(1+z')^3  {\vec
B_0}({\hat q}r') \cdot {\hat q} { {\rm d} r' \over {\sqrt {1 - Kr^{'2}}}
} ={\RM ({\hat q}, z) \over \nu_0^2} \, .
\end{equation}
Today's values are all denoted by the subscript ``0" and $z'=z(r')$.
The time dependence of $\delta({\hat q}r')$ is taken into account later
on (\cf Eq. \ref{eq:ups_xi}).
The two point correlation function of the RM is defined by
\begin{equation}
\label{eq:ups_def}
\Upsilon \equiv \langle \RM ({\hat q}_1, z_1) \RM ({\hat q}_2, z_2)
\rangle \,, 
\end{equation}
see Nissen \& Thielheim (1975) for a similar definition.
Throughout the paper, $\langle ... \rangle$ is the notation for an
ensemble average.
We use the notation $\vec r\,' =
\vec r_1\>' - \vec r_2\>' $ and obtain 
\[
 \Upsilon =  \left( {e^3 \bar n_{b_0} \over 2
\pi m_e^2 } \right)^2 \int_0^{r_1(z_1)}
{ {\rm d} r'_1 \over {\sqrt {1 - Kr_1^{'2}}} }
\left[1+\delta(\hat q_1 r_1')\right](1+z'_1)^3 
\int_0^{r_2(z_2)}
{ {\rm d} r'_2 \over {\sqrt {1 - Kr_2^{'2}}} }
\left[1+\delta(\hat{q}_2r_2')\right] (1+z'_2)^3
\]
\begin{equation}
\times
\label{eq:ups_1}
\hat {q}_{1i} \hat {q}_{2j} {1 \over V} \int B_{0i}({\vec x})
B_{0j}({\vec x} + \vec r\>') {\rm d}^3 {\vec x}
\,, 
\end{equation}
where summation over the spatial components $i,j$ is assumed.
The integral (\ref{eq:ups_1}) is made out of four terms. Two of the
terms (those involve only one power of $\delta$) vanish because $\langle
\delta \rangle = 0$ and we assume vanishing correlation between the density
fluctuations and magnetic fluctuations [$\langle \vec B(\vec x) \delta(\vec x +
\vec r) \rangle = 0\, \forall (\vec x,\vec r)$] due to the vector nature
of $\vec B$ and the scalar $\delta$. 

The two remaining terms are
\[
 \Upsilon_0 =  \left( {e^3 \bar n_{b_0} \over 2
\pi m_e^2 } \right)^2 \int_0^{r_1(z_1)}
{ {\rm d} r'_1 \over {\sqrt {1 - Kr_1^{'2}}} }
(1+z'_1)^3 
\int_0^{r_2(z_2)} 
{ {\rm d} r'_2 \over {\sqrt {1 - Kr_2^{'2}}} }
(1+z'_2)^3
\times  
\]
\begin{equation}
\label{eq:ups_0}
{\hat q}_{1i} {\hat q}_{2j} {1 \over V} \int B_{0i}({\vec x})
B_{0j}({\vec x} + \vec r\>') {\rm d}^3 {\vec x}
\,, 
\end{equation}
and
\[
 \Upsilon_\xi =  \left( {e^3 \bar n_{b_0} \over 2
\pi m_e^2  } \right)^2 \int_0^{r_1(z_1)}
{ {\rm d} r'_1 \over {\sqrt {1 - Kr_1^{'2}}} }
(1+z'_1)^2 
\int_0^{r_2(z_2)} 
{ {\rm d} r'_2 \over {\sqrt {1 - Kr_2^{'2}}} } 
\xi_0( \vec r\>' ) (1+z'_2)^2
\times  
\]
\begin{equation}
\label{eq:ups_xi}
{\hat q}_{1i} {\hat q}_{2j} {1 \over V} \int B_{0i}({\vec x})
B_{0j}({\vec x} + \vec r\>') {\rm d}^3 {\vec x}
\,, 
\end{equation}
where $\xi(r)$ is the baryonic matter correlation function that we identify
with the matter correlation function. Today's value of it is $\xi_0$ and
the missing powers of $(1+z)$ in Eq. (\ref{eq:ups_xi}) account for
the linear evolution of this correlation. The RM correlation function is
the sum $\Upsilon = \Upsilon_0 + \Upsilon_\xi$.

\begin{figure}[t!]
\vskip-1.2truecm
\hskip5truecm {\epsfxsize=2.7 in \epsfbox{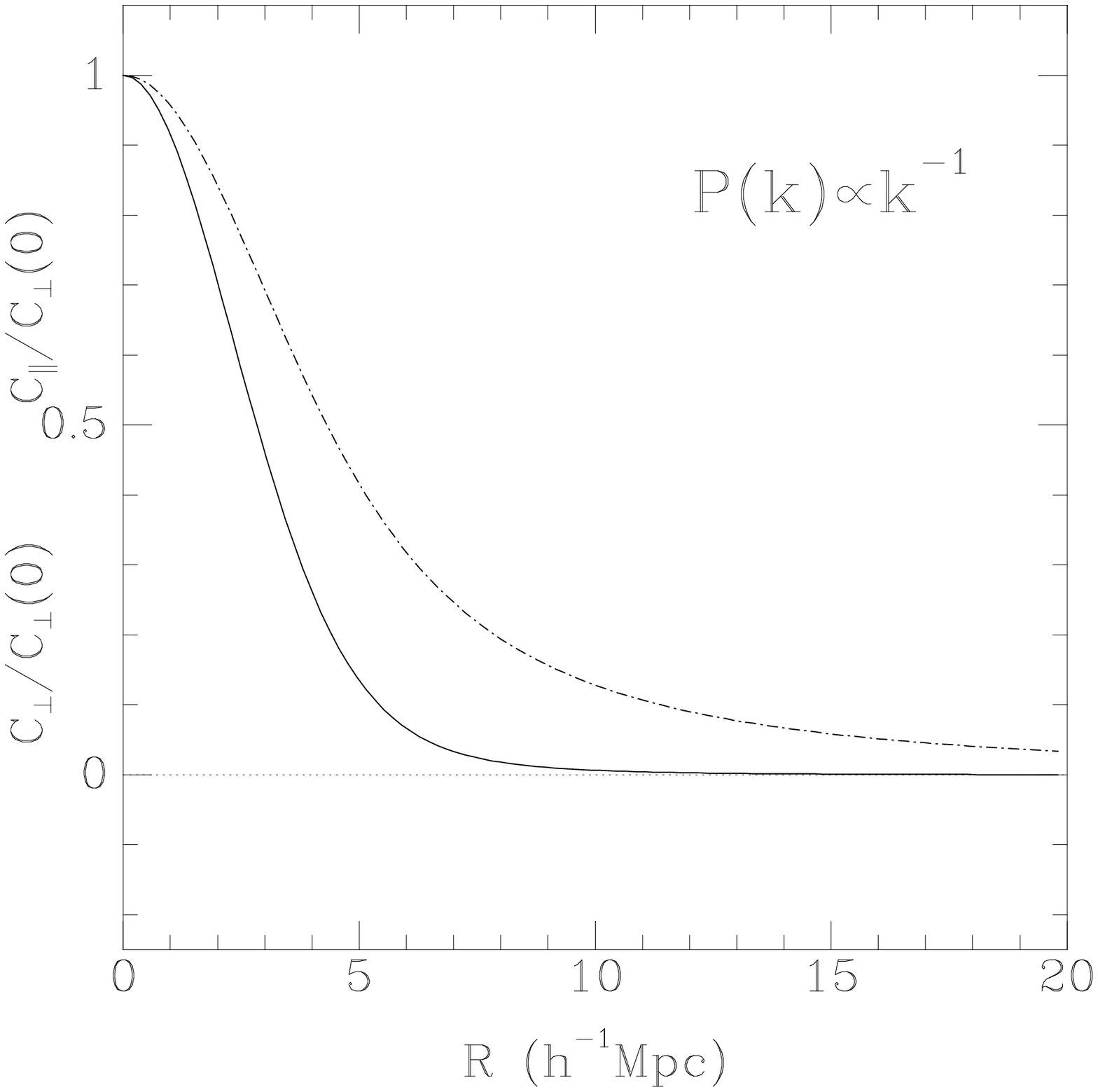}}
\vskip-2truecm
\hskip0.5truecm {\epsfxsize=2.7 in \epsfbox{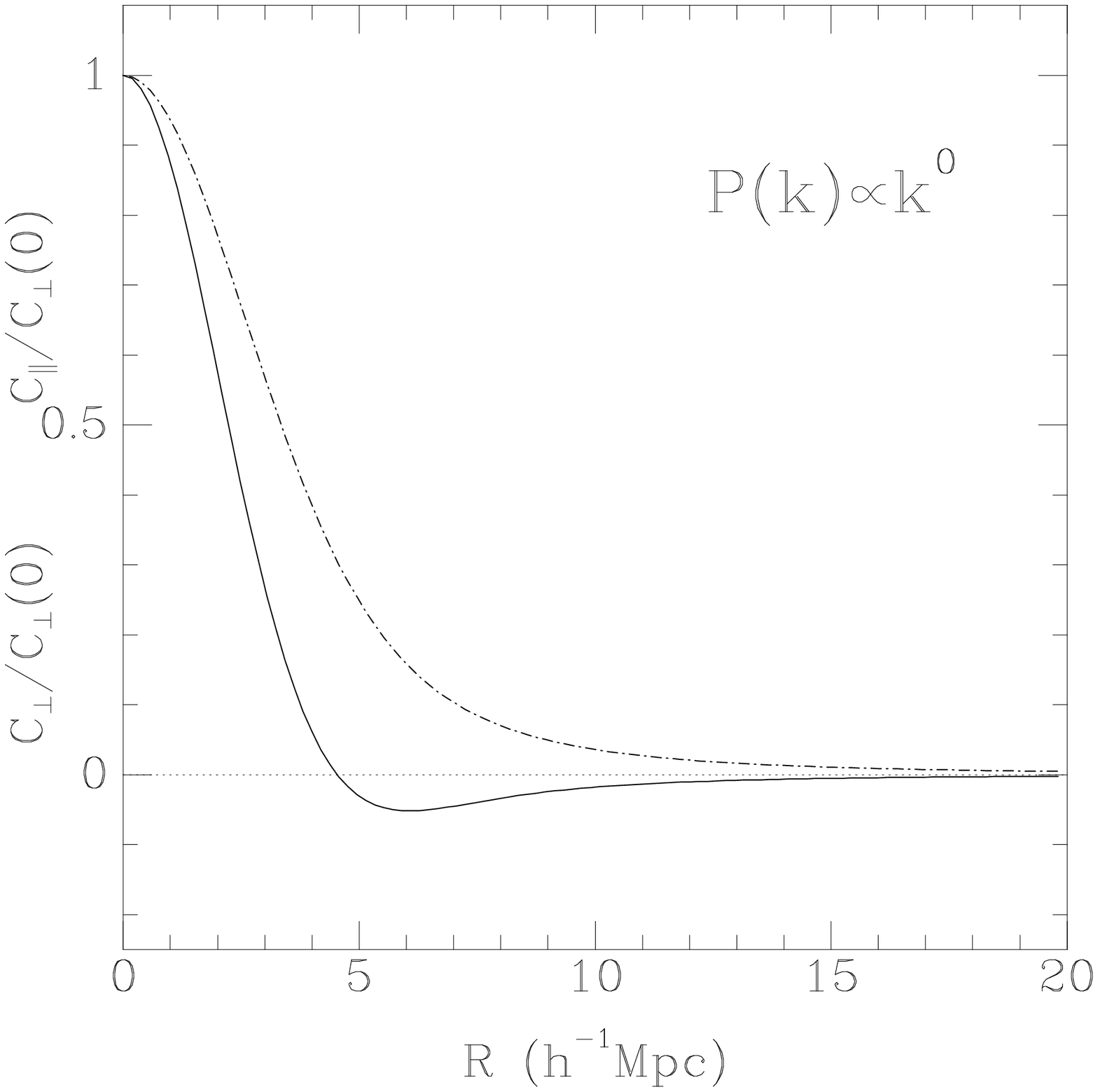}}
\hskip2truecm {\epsfxsize=2.7 in \epsfbox{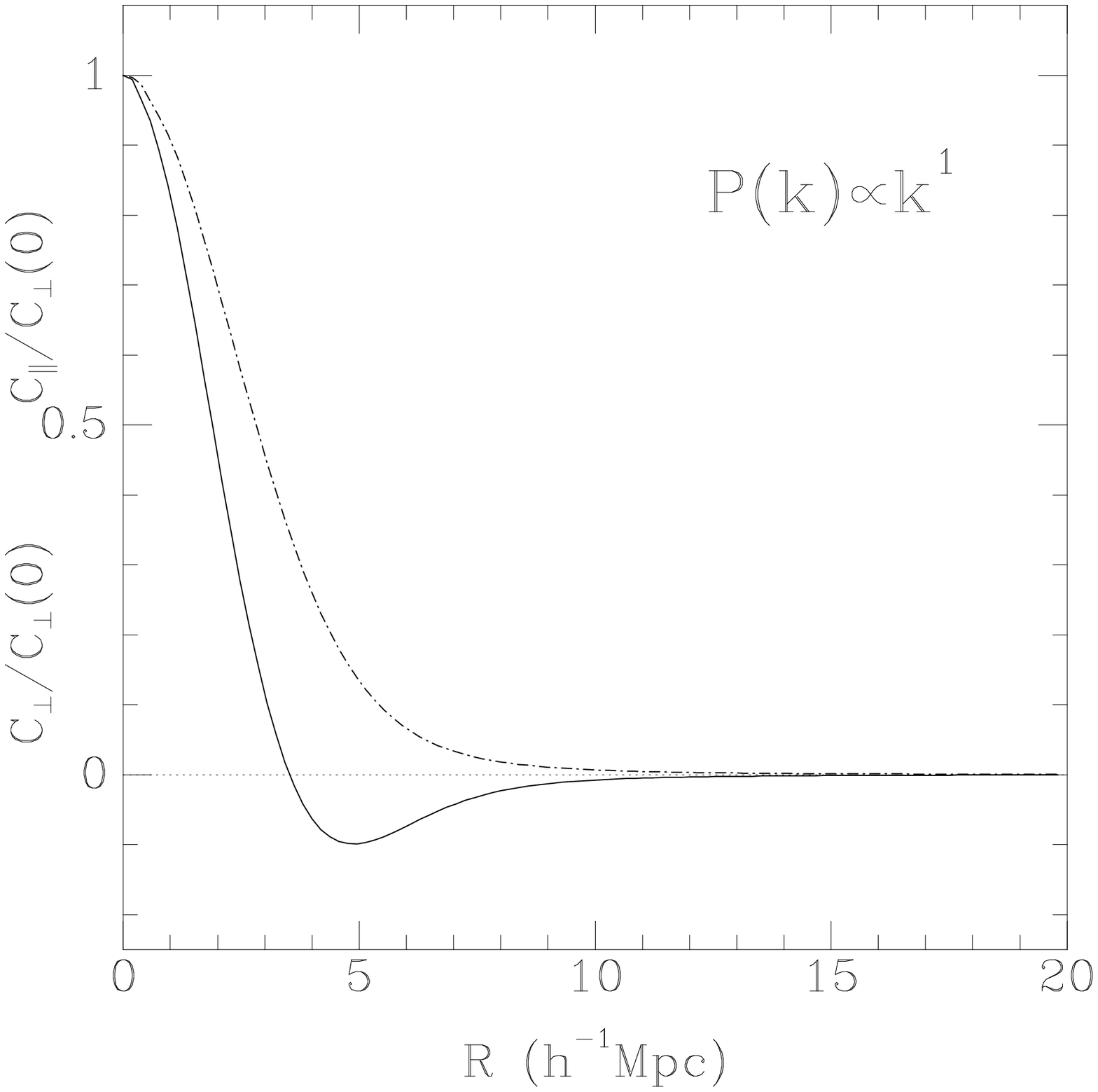}}
\vskip-0truecm
\caption{\baselineskip 0.4cm{\trm
The parallel (dotted-dashed line) and perpendicular (solid line)
components of the magnetic field
correlation tensor. Units are of $C_\perp (=C_\parallel)$  in the
origin.
Plotted are three cases: (a) $P_B(k) \propto k^{-1}$, (b) $P_B(k) = const.$,
(c) $P_B(k) \propto k$. Gaussian smoothing of $1.5\hmpc$  applied.
}}
\label{fig:b_corr}
\end{figure}

The last integral of Eqs. (\ref{eq:ups_1}, \ref{eq:ups_0},
\ref{eq:ups_xi}) is the correlation tensor of the magnetic field
defined as $C_{ij}({\vec r}) = \langle B_i({\vec x}) B_j ({\vec x}+{\vec r}) \rangle$. In the
cosmological case, we assume isotropy on scales much smaller than the
horizon and confine $C_{ij}$ to be a function of $\vert {\vec r} \vert$ only.
Being a divergence-free field the magnetic field correlation can be
written (Monin \& Yaglom 1975) as a combination of parallel
($C_{\parallel}$)
and perpendicular ($C_{\perp}$)
functions. The parallelism and orthogonality are given with respect to the
connecting vector:  ${\vec r} = {\vec r}_1 - {\vec r}_2$
\[
C_{ij}({\vec r}) = [C_{\parallel}(r)- C_{\perp}(r)]{r_i r_j \over r^2} +
C_{\perp}(r) \delta^K_{ij}\, , 
\]
\[
C_{jl}({\vec k}) = \int C_{jl}(\vec r) \, \exp(-i\vec k \cdot
\vec r) \, d^3r
\]
\begin{equation}
C_{jl}({\vec k}) = [C_{\parallel}(k) - C_{\perp}(k)]{k_j k_l \over k^2}
+ C_{\perp}(k) \delta^K_{jl}\, ,
\label{eq:c_pre_par}
\end{equation}
where $\delta^K_{ij}$ is the Kronecker $\delta$-function.
In the spectral domain we define the function $E(k)=4 \pi k^2
C_{\perp}(k)$ and express both correlation functions by
the function $E(k)$, namely
\begin{equation}
\label{eq:c_pre_par_2}
C_{\perp}(r) = \int_0^\infty {\rm d} k E(k) \left(j_0(kr) - {j_1(kr) \over kr}\right) \quad ; \quad
C_{\parallel}(r) = 2 \int_0^\infty {\rm d} k E(k) {j_1(kr) \over kr}\, , 
\end{equation}
with $j_i$ denoting the spherical Bessel function of the $i^{th}$ order.
When people refer to the ``magnetic power spectrum" they usually mean
$C_{\perp}(k)$ which should further be multiplied by $k^2$ in order to
get $E(k)$. We shall hereafter comply with this notation and refer to
$C_{\perp}(k)$ as the three-dimensional ``magnetic power spectrum", $P_B(k)$.
Figure ~\ref{fig:b_corr} show the two functions $C_{\parallel}(r)$ and
$C_{\perp}(r)$ in units of $C_{\perp}(0)$ for three values of the power
index $n=-1,0,1$ [$P_B(k) \propto k^n$]. Each function is shown with a minimal Gaussian smoothing 
of $1.5 \hmpc$.
Naturally, smaller power indices mean larger correlation length.

Going back to the expression (\ref{eq:ups_1}) for the RM correlation we can rewrite it as
\[
\Upsilon_0 =  \left( {e^3 \bar n_{b_0} \over 2
\pi m_e^2 } \right)^2 \int_0^{r_1(z_1)}
{ {\rm d} r'_1 \over {\sqrt {1 - Kr_1^{'2}}} }
(1+z'_1)^3
\int_0^{r_2(z_2)} 
{ {\rm d} r'_2 \over {\sqrt {1 - Kr_2^{'2}}} }
(1+z'_2)^3
\times \]
\begin{equation}
\label{eq:ups_2}
\left\{ [C_{\parallel}(r')- C_{\perp}(r')]({\hat {q}_1 \cdot \hat r')
(\hat{q}_2 \cdot \hat r')} +
C_{\perp}(r'){\hat q}_1 \cdot {\hat q}_2 \right\}
\,. 
\end{equation}
As expected, the RM correlation depends only on the relative distance,
$ \vert {\vec r} \vert$,
and the angle.

For comoving distances between two sources at $z_1$ and $z_2$
separated by the angle $\gamma$ on the sky we use Weinberg's (1972,
Eq. 14.2.7) coordinate transformation, implemented by Osmer (1981) for
the $q_0=0$ case ($q_0$ -- the deceleration parameter) and generalized by Boyle (1986) for other $q_0$
values. For the flat universe case ($q_0=0.5$) the comoving relative
distance reduces to the simple form
\begin{equation}
\label{eq:r1_r2_z}
r(z_1, z_2, \gamma) = {1 \over H_0} \left[r_1(z_1)^2+r_2(z_2)^2-2r_1(z_1)r_2(z_2)\cos\gamma\right]^{1/2}\, . 
\end{equation}
The distance to a source at the redshift $z$ (no cosmological constant) is given by 
(Kolb \& Turner 1990, Eq. 3.112)
\begin{equation}
\label{eq:r_of_z}
r(z) = {1 \over  H_0} {2\Omega_0z + (2\Omega_0-4)({\sqrt {\Omega_0z+1}}-1) \over \Omega_0^2(1+z) } \,,  
\end{equation}
and can be easily generalized for other cases.

In order to estimate the importance of $\Upsilon_\xi$ [Eq.
(\ref{eq:ups_xi})] we adopt the
APM power spectrum (Baugh \& Efstathiou 1993; Tadros \& Efstathiou 1995)
with the fitting formula
\begin{equation}
P_m(k) =
{2\pi^2 \over k^3}\, { (k/k_0)^{3-m} \over 1+ (k/k_c)^{-(m+n)} } \,,
\label{eq:p_m}
\end{equation}
the fitting parameters $m=1.4$, $k_c=0.02$, $k_0=0.19$, and Gaussian smoothing
of $1.5 \hmpc$. We then derive the real space correlation function
by the Fourier transform of the smoothed power spectrum. In the range $3-30
\hmpc$ the result is very similar to the derived APM real space
correlation function (no explicit smoothing) 
$\xi_0=(r/5.25 \hmpc)^{-1.7}$ (Baugh 1995).
For this choice
we get (regardless of the $P_B(k)$ form, and for $z_{src} \simgreat 0.5$), $\vert \Upsilon_\xi/\Upsilon \vert
< 10^{-4}$. We hence identify $\Upsilon = \Upsilon_0$ hereafter.

\subsection{The Power Spectrum Normalization}
\label{subsec:norm}
We relate the magnetic field normalization to the power spectrum by 
$B_{rms}(R)$ where
\begin{equation}
\label{eq:b_norm}
(B^{G,TH}_{rms})^2(R) = {3 \over 2\pi^2} \int_0^\infty P_B(k) k^2 W_{G,TH}^2(kR) {\rm d} k \, ,  
\end{equation}
the factor $3$ is due to the definition of $P_B(k)$ [as
$C_{\perp}(r)$], and $W_{G,TH}(kR)$ is 
the Gaussian (Top-hat) window function of radius $R$, in $k$ space,
\begin{equation}
\label{eq:w_g}
W_G = \exp \left[ - {(kR)^2 \over 2} \right] \, ; \quad
W_{TH} = { 3 \over (kR)^3 } [ \sin(kR) - kR\cos(kR)] \,.
\end{equation}
A good benchmark to use for the normalization range is the 
CMB limit due to its simplicity and the availability of 
$\Delta T /T$ measurements on many scales (either today or in the near future).
We therefore work in the range of $\sim 1$ nG throughout this paper.

The observational limits on the magnetic field as deduced from the
$\RM$ dipole [$R\simeq r(z=2.5)$ or $R\simeq r(z=3.6)$] are thus very different from the limits imposed by CMB
fluctuations on $1'$ scale. We note that by Eq. (\ref{eq:b_norm}) one cannot infer
limits on the magnetic field magnitude from one scale to another. The
limits must involve the power spectrum shape. Since no estimate for the
latter exists, and since theoretical predictions for it range from
power index of $n=-3$ to $n=2$ (Ratra 1992a), limits on one particular
scale hardly limit other scales.

\begin{figure}[t!]
\vskip-1.5truecm
\hskip5truecm {\epsfxsize=2.7 in \epsfbox{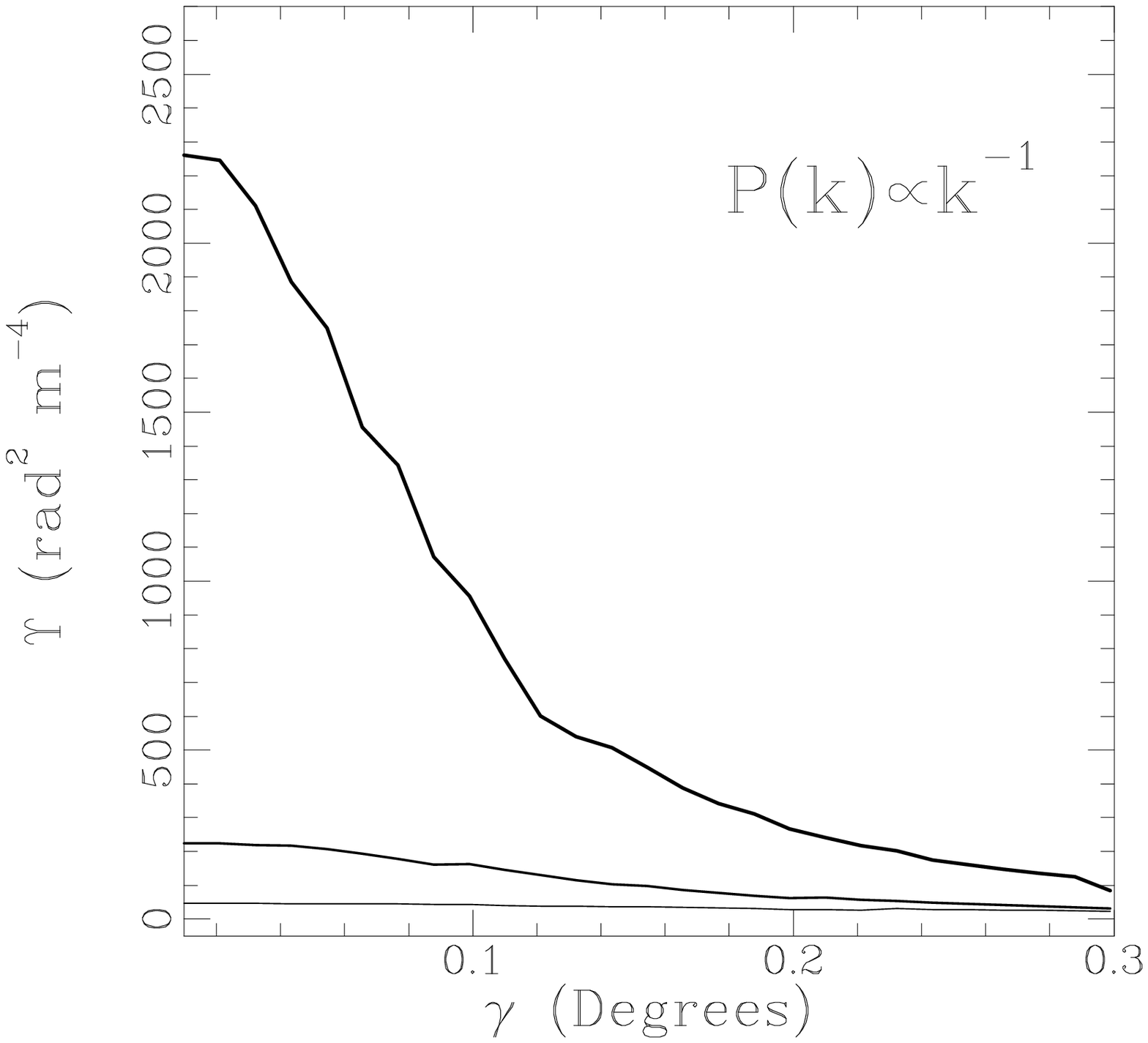}}
\vskip-3truecm
\hskip0.5truecm {\epsfxsize=2.7 in \epsfbox{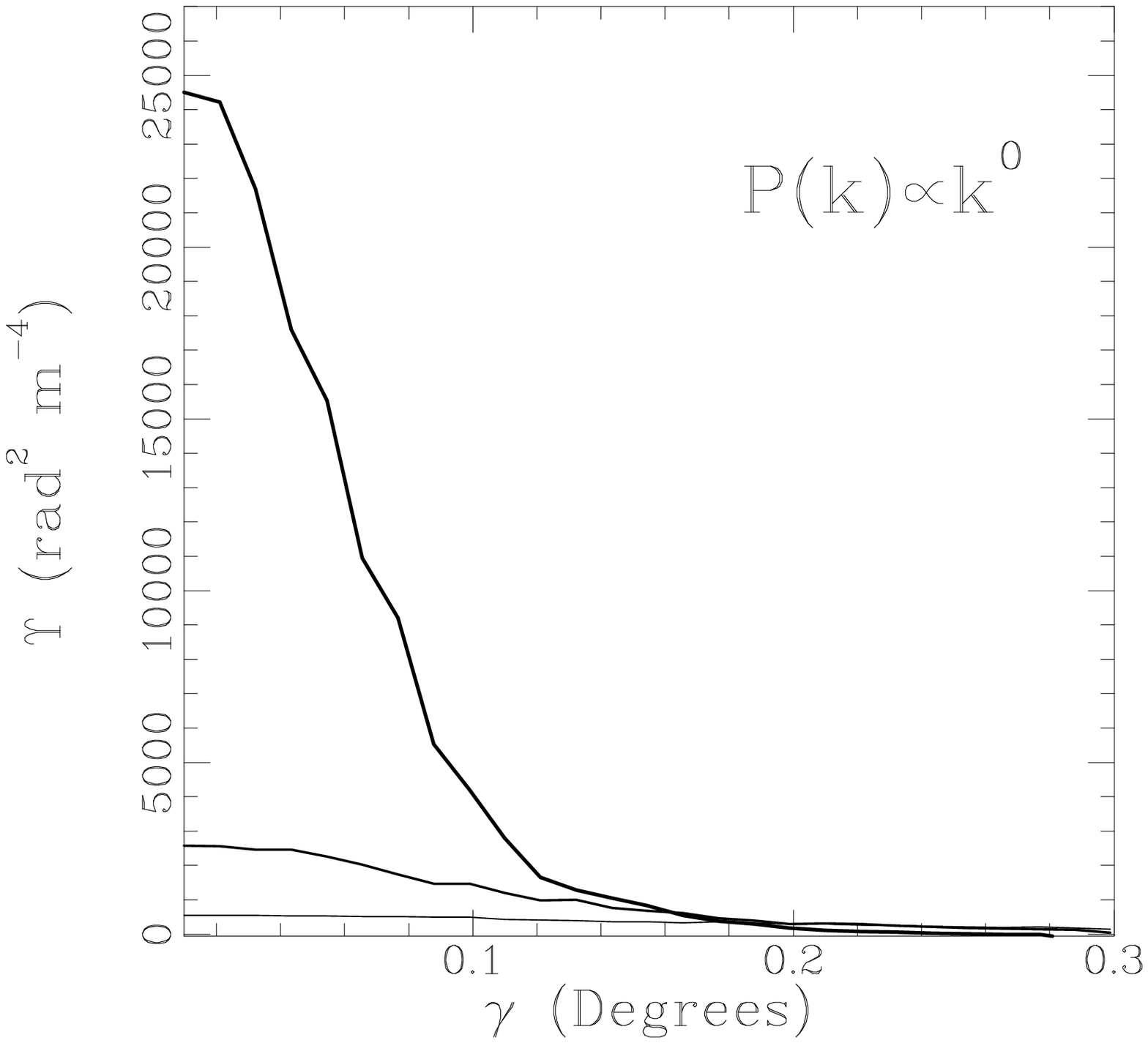}}
\hskip2truecm {\epsfxsize=2.7 in \epsfbox{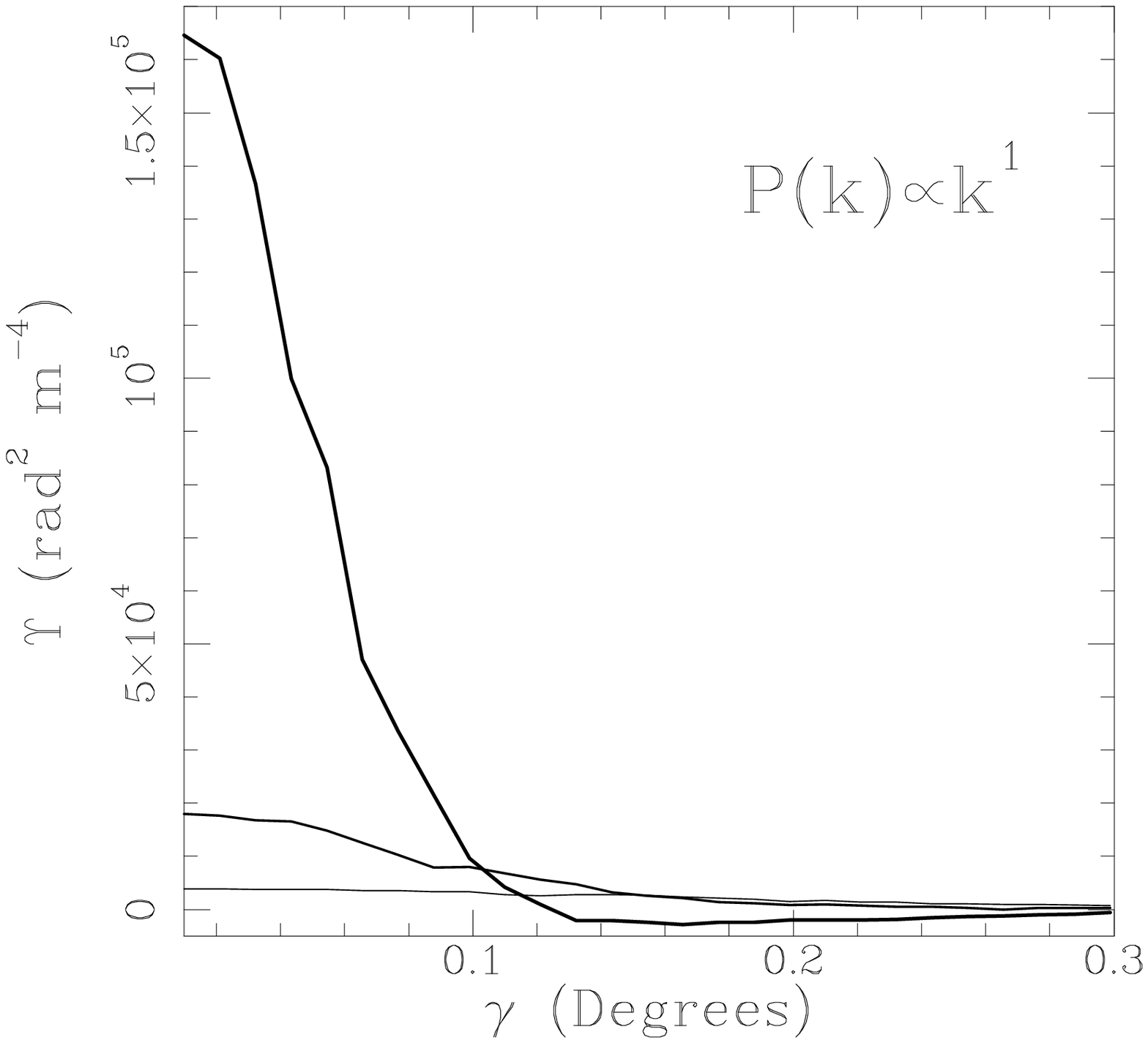}}
\vskip-1truecm
\caption{\baselineskip 0.4cm{\trm
RM correlation as function of the separation angle between source pairs
at three different redshifts. The redshift values are $z_{src}=0.5,1,2$
with thicker lines for higher redshift. The magnetic field normalization
in all cases is $B_{rms}^{TH}(50\hmpc)=1$ nG .
Three magnetic power spectra are shown with the
power indices $n=-1,0,1$. Note the different $y$ axis scales.
}}
\label{fig:rm_corr}
\end{figure}

Figure \ref{fig:rm_corr} show the values for the RM correlation function
for pairs of sources at the same redshift, and three different spectral
indices. 
The normalization of the magnetic field in each case is $B_{rms}^{TH}(50\hmpc) =
1$ nG ($\Omega_m=1, h=1, \Omega_b = 0.024$). This scale resembles the
mean separation between rich clusters.

We notice that for close pairs, high redshift sources are
preferable in order to get high signal, while a transition typically occurs at
a fraction of a degree where it becomes preferable to use lower
redshift sources. This transition is due to the uncorrelated magnetic fields
experienced by high redshift sources with large separation. Only when
the two lines-of-sight approach the correlation length, does contribution to
the RM correlation becomes substantial.
The different, uncorrelated magnetic fields up to the correlation
length, play the role of uncorrelated noise, and dominate the
correlated signal from the low redshift segment of the line-of-sight.

On the scale $0.5^\circ - 2.5^\circ$, the RM correlation signal
still stands on $\sim 10-10^2 \,{\rm rad}^2 \, {\rm m}^{-4}$ depending on the
power spectrum.

\section{CALCULATION PROCEDURE FOR THE CORRELATION MATRIX}
\label{sec:calc}

\subsection{The Raw Data }
\label{subsec:raw}
The raw data consist of $N_{src}$ extra Galactic objects for which
polarization measurements are available in $N_\lambda$ wavelengths.
Broten, Macleod, \& Vall\'ee (1988), and Oren \& Wolfe  (1995) have 
emphasized the need for a careful
selection of wavelengths for each observed source. For the current
application there are additional special requirements that will become
clear when we get to the discussion (\S\ref{sec:discuss}).\\
Each of such
$N_\lambda$ measurements consists of two types of data : the degree of
polarization, and the polarization position angle [$\vert p \vert$ and
$2\phi$ of Eq. (\ref{eq:p_vec})]. Two functions
are then constructed which give the dependence of these on wavelength. The
$\RM$ is computed from the latter by using Eq. (\ref{eq:p_vec})
where $\RM = \phi\, \lambda^{-2}$.
Each measured $\RM_m$ value has an error, $\epsilon_m$,
that emerges from the measurement error (instrumental, ionosphere model, the
earth magnetic field model etc.) 
and translates to the error in the fit from which the $\RM$ is derived.
For typical sources with polarization degree of \gtsima $10\%$
(namely $10^{-1}-10^{-2}\,{\rm Jy}$ of polarized component at a few GHz
frequency), the measurement
error in each datum is of the order of a few degrees (\eg Kato, Tabara,
Inoue \& Aizu 1987; Simard-Normandin, Kronberg, Button, 1980 (SKB); Oren
\& Wolfe 1995). 
For realistic $N_\lambda=4$, this instrumental
error translates to a typical error in the $\phi - \lambda^{2}$ fit of
$0.5-5\radm2$ and is dwarfed
by other noise terms in the procedure (\cf \S\ref{subsec:gal_mask},
\S\ref{subsec:I+f}). We therefore set $\epsilon_m
\simeq 0$ from now on.

The measured RMs are the sum of a few
contributions, and cannot be directly plugged into the expression (\ref{eq:ups_2}) to
evaluate $\Upsilon$.\\
To begin with, there exists the internal $\RM_{I}$ for every
one of the $N_{src}$  extragalactic sources. Then there is the integrated
$\RM_c$ which is presumably due to the cosmological magnetic field (and
the free electrons) along the line-of-sight to the source. 
There may also exist contribution $\RM_f$ from intervening systems along the
line-of-sight (foreground screen) either next to the source itself
(\eg the host galaxy of a quasar), Lyman-$\alpha$ systems,
galaxies, or clusters of galaxies.
Before this combined signal gets to the detector it still has to go 
through the Galactic magnetic field, where it is rotated once more by the amount
$\RM_G$. The final measurement, $\RM_m$ is thus the linear sum
\begin{equation}
\label{eq:lin_sum}
RM_m({\vec z}_i)=\RM_I^i + \RM_c({\vec z}_i) +  \RM_f({\vec z}_i) + 
\RM_G(\hat z_i) \, ,
\end{equation}
where ${\vec z}_i$ is the redshift-vector [actually translated to a
distance vector (\cf \S2)] for the $i^{th}$ source.
Since we are interested in the
cosmological contribution to the $\RM$ [\ie the second term in Eq.
(\ref{eq:lin_sum})],
the first step should involve ``cleaning" the measured signal
of all irrelevant extra contributions. We shall attempt to perform this cleaning
in a statistical way.

\subsection{The Galactic Mask}
\label{subsec:gal_mask}
There are two alternatives to assess the Galactic contribution to
the measured $\RM$. The two ways differ by the population sample used
for the assessment. If an independent population at the outskirts of the
Galaxy exists, for which $\RM$ can be measured, this population can be used
to map out the Galactic $\RM$. We shall hereafter use the term ``mask
source population'' for the set of objects by which we map the Galactic
contribution to the RM.

An estimate for the thickness of the Galactic magnetic layer is
$\sim 1$ kpc (Simard-Normandin \& Kronberg 1980). 
That means that apart from the Galactic center direction, sources of
distances that satisfy $r_{src}>1/\sin(\vert b \vert )$ kpc ($\sim 3$
kpc for $\vert b \vert = 20^\circ$) are located
beyond the Reynolds layer and fully probe the Galactic contribution
to the RM. We hence consider two possibilities for the mask source
population.

One natural candidate for this role is
the pulsar population, for which $\RM$s are available.
In order to make use of the pulsars we need to find a subset of them
that reside in the outer part of the Milky-way magnetic layer.\\
A class of pulsars that is especially appropriate for the task of
probing the Galactic RM contribution is the millisecond pulsars. This
population of old pulsars is believed to reside far out of the Galactic
plane. The number density of millisecond pulsars as implied by a number of surveys 
at high Galactic latitude ($\sim1\, {\rm mJy}$ sensitivity at $\sim1$ GHz) ranges between $0.01 - 0.0175$ deg$^{-2}$, \ie
$400 - 700$ pulsars over the sky (Foster, Cadwell, Wolszczan, \&
Anderson 1995; Camilo, Nice, \& Taylor 1996).

The alternative mask population for assessment of the Galactic $\RM$ 
contribution is a sub-set of the closest extragalactic
sources. The disadvantage of using this population is the need
for a large enough source number at low enough redshift, to allow
differentiation between the cosmological contribution and the Galactic one.
Exploiting extragalactic sources for the resolution of the Galactic
contribution affects only slightly the amount of cosmological contribution
correlation, as long as the nearby sources are taken within $z\simless 0.1$, 
since the bulk of the sources is at $z \simgreat 1$ (\cf \S\ref{sec:testing}).

In order to use either one of the galactic mask populations, its number
density should exceed a certain minimal number density. We now attempt
to assess this minimal number.\\
The Galactic $\RM$ contribution is a two-dimensional field with $\RM$
values. The data on the other hand are made of distinctive sources
located at discrete directions. The lines-of-sight to the Galactic mask
population, do not necessarily coincide with the
line-of-sight direction towards the extragalactic sources. The way to
circumvent this difficulty is by introducing a {\it smoothed} version of
the Galactic $\RM$ map, namely the field  $\rmgs$. The two fields are
connected via the smoothing angle $\theta_s$, and by a specific choice of
a Gaussian smoothing
for $N_G$ measurements of the Galactic $\RM_G(\hat z_i)$ located at the
directions $\hat z_i$
\begin{equation}
\label{eq:gal_sm}
\rmgs = 
{\sum\limits_{i=1}^{N_G} \RM_G^i \,\exp \left[- { ({\hat z}_i - \hat{z})^2 \over
2 \theta_s } \right] \over \sum\limits_{i=1}^{N_G} \exp \left[- { ({\hat z}_i
- \hat{z})^2 \over
2 \theta_s^2 } \right] } \,.
\end{equation}
When implemented using real data, the smoothing involves weights by
measurement errors as well. 
For the purpose of this paper, we ignore these weights
to avoid unnecessary complication.
Using a smoothed field for the Galactic $\RM$ contribution, introduces
random noise, $\epsilon_G$, that can be directly evaluated by the very same
$N_G$ sample via
\begin{equation}
\label{eq:gal_res}
\epsilon_G^2 = {1 \over N_G} \sum_{i=1}^{n_G} \,
\left(\RM_G( {\hat {z}_i}) -{\widehat{\RM}_G}( {\hat {z}_i}) \right)^2 \,.
\end{equation}

\begin{figure}[t!]
\plotfiddle{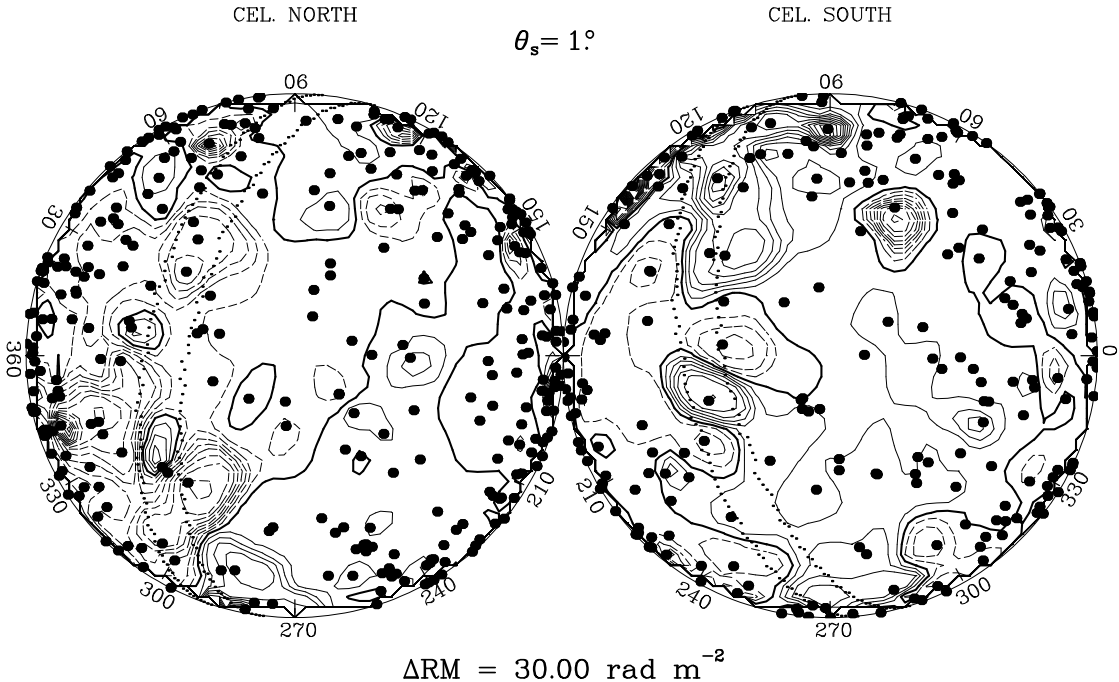}{5truecm}{270}{65}{65}{-264}{200}
\vskip4truecm
\caption{\baselineskip 0.4cm{\trm
The $1^\circ$ smoothed RM Galactic mask projected onto celestial
coordinates.
Positive RM values marked by
continuous lines and negative values by dashed lines. The $\RM=0$ line
is thicker. The spacing is
$\Delta\RM = 30 \radm2$. The locations of sources in the
SKB catalog (projected) are marked by solid dots,
and the Galactic plane is marked by the dotted strip.
}}
\label{fig:maskmap}
\vskip0.5truecm
\end{figure}

\begin{figure}[t!]
\plotfiddle{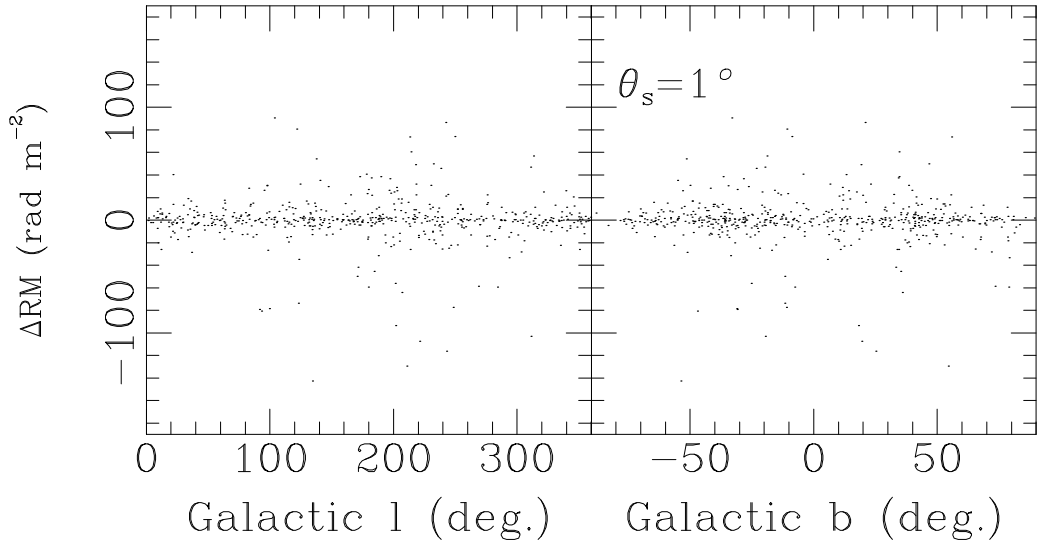}{5truecm}{270}{65}{65}{-264}{300}
\vskip1.0truecm
\caption{\baselineskip 0.4cm{\trm
The RM residuals from the $1^\circ$ smoothed Galactic mask
as function of the Galactic longitude (left) and latitude (right).
}}
\label{fig:res_mask}
\end{figure}

\begin{figure}[t!]
\plotfiddle{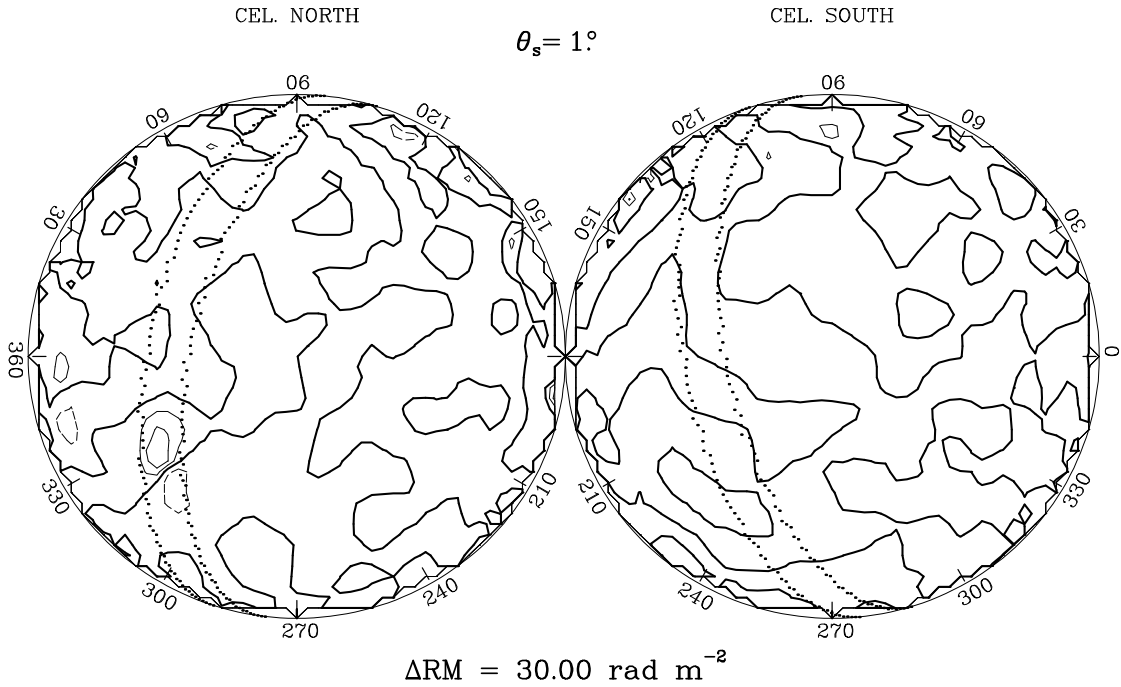}{5truecm}{270}{65}{65}{-264}{200}
\vskip4truecm
\caption{\baselineskip 0.4cm{\trm
The $1^\circ$ smoothed RM residuals of the Galactic mask in celestial
coordinates.
Symbols as in figure (\ref{fig:maskmap}).
}}
\label{fig:maskres_map}
\end{figure}

The noise can be a function of position, it is difficult however to
account for a position-dependent noise with low number of $N_G$. 
There are a few estimates in the literature for the noise level, $\epsilon_G$,
on different smoothing scales.\\
An upper limit for $\epsilon_G$ is given in Minter \& Spangler (1996), who
calculate the observed ``structure function'', $D_{\rm RM}$ which is the
same as Eq. (\ref{eq:gal_res}) but with an arbitrary lag $\gamma$ and
the unsmoothed values of the RM. 
They conclude that $  D_{\rm RM}(\gamma>1^\circ)
=(340\pm30)\gamma^{0.64\pm0.06}$rad$^{2}$m$^{-4}$, and drops steeper for
$\gamma<1^\circ$.
The value of the $D_{\rm RM}$ is an upper limit for $\epsilon_G^2(\gamma)$
as calculated here, due to the smoothing we apply (and was not
applied in the $D_{\rm RM}$ calculation). Simonetti, Cordes, \&
Spangler (1984) calculate $D_{\rm RM} = 484-8100\,$ rad$^2$m$^{-4}$,
depending on the Galactic latitude, for much
larger angular scales of $30^\circ-50^\circ$ ! (linear scale).\\
Simonetti \& Cordes (1986) corroborate these results and consider even
larger angular scales to obtain typically $D_{\rm RM}<1000\,$rad$^2$m$^{-4}$
on scales less than $100^\circ$.\\
Oren \& Wolfe (1995) calculate a very similar
quantity to $\epsilon_G$ (with a varying top hat window instead of a
fixed Gaussian), and obtain a typical variance of $<1000\,$ rad$^2$m$^{-4}$ on
$\sim 30^\circ$ scales.\\
All of these measurements suggest that for a mask population with number
density of about one source per $1000\,$deg.$^2$, the Galactic
contribution to the RM ($\vert b \vert > 20^\circ$) can be resolved with a $1\sigma$ accuracy of $30
\radm2$. That means that an isotropic coverage of about $100-200$ mask
sources on the sky is enough for achieving this noise level.\\
``Damage control"  for ignoring the position dependence can be devised
by comparing the ``clean" RM field dipole and quadrupole to the Galactic
direction of the two. These first two multipole should be the most
prominent signature of the Galactic RM mask.

We go back to the suggested mask populations and examine whether they
are able to fulfill the requirement of the minimal number density.\\
The largest pulsar catalog to date [the electronic version of
Taylor, Manchester, \& Lyne (1993)] consists of $800$ pulsars out of
which $259$ have RM measurements. This catalog was not compiled with a
specific emphasis for millisecond pulsars, and therefore only $124$
pulsars with RM values lie in Galactic latitude $\vert b \vert>5^\circ$.
The softer condition $r_{src}> 2/\sin(\vert b \vert)$ (\ie considering an ``RM 
layer" of $4$ kpc) is fulfilled for $237$ pulsars with both RM registered
value and estimated distance $r_{src}$ (Taylor \& Cordes 1993).\\
However, these $237$ pulsars are still not distributed isotropically
across the sky, and tend to concentrate near the Galactic plane.

A way to avoid this anisotropy is to compile a sample of millisecond
pulsars, that lie far away from the Galactic plane.
This population has a number density that exceeds the necessary
$\sim10^{-2}$ deg.$^{-2}$ by at least a factor of two and it may be as high as seven
times denser than the minimal value.
This population can be detected for RM (Fernando Camilo, private
communication, 1997)
\footnote{In the Taylor, Manchester, \& Lyne catalog there are two
pulsars with both RM registered value and $p<10$ ms, 
and five for which $p<100$ ms.
The smallest polarized flux at $1.4\,$GHz for a pulsar with a registered
RM is $0.2\,$mJy}
and ensures the $\epsilon_G \simeq 30\radm2$ noise level.
This kind of sample is ideal for the current paper analysis, as it does
not contain any cosmological contribution or internal RM contribution.
The main disadvantage of it is that it does not exist yet.

In considering the extragalactic mask population;
for the $\simless 400$ sources needed to resolve the 
Galactic contribution, it is sufficient to have a sample of $1$mJy 
sensitivity (at a few GHz) up to $z\simeq0.1$ 
[see Loan, Wall, \& Lahav (1997) for the
number density of radio sources with flux $ > 35$ mJy, and Dunlop \&
Peacock (1990) for the combined flat/steep spectrum luminosity function
for extrapolation to the $1$mJy limit].  

For this paper we follow Oren \& Wolfe (1995) and use an extragalactic 
sample of RM measurements as an
upper limit for the noise that appears by the Galactic RM removal
procedure. The sample we use consists of $555$ sources as listed by
SKB. 
The SKB catalog allows us to
realistically mimic the Galactic mask, and is an overestimate since we
ascribe all the RM of the sources to the Galactic contribution. Moreover,
unlike Oren \& Wolfe we do not exclude ``outliers". This use aims to
provide a realistic model of the Galactic mask contribution, and allows
us to demonstrate the method power. For our purpose it should not be taken 
literally as the true Galactic mask contribution 
(even though Oren \& Wolfe did consider it that way) because no
segregation by redshift was applied to choose mask sources from the SKB
list.
We neither attempted to evaluate the internal RM (by means explained
in \S\ref{subsec:I+f}) in order to minimize the $\epsilon_G$ value.\\

The SKB source locations (projected, and therefore concentrated toward the
circumference) are marked as black dots on Figure \ref{fig:maskmap}
and shown on top of the smoothed Galactic $\RM$ field with smoothing scale
of $\theta_s=1^\circ$. The smooth map is shown in celestial coordinates.
Figure ~\ref{fig:res_mask} shows the RM residuals as functions of
Galactic latitude and longitude, and figure ~\ref{fig:maskres_map}
shows the contour map of the smoothed residual field.
Both figures (\ref{fig:res_mask} \& \ref{fig:maskres_map}) show hardly any longitude dependence of the residuals (in
spite of the solar system off-center position), and bigger residuals in
the $\sim\pm20^\circ$ strip about the Galactic plane (\cf the same
conclusion of Oren \& Wolfe 1995).
In an all-sky coverage of sources, this residual deviation should be of 
some worry, but as we shall see (\S\ref{subsec:all_sky}), it doesn't bias the result for the most 
likely power 
spectrum as calculated by the Bayesian analysis. If the analysis is confined 
to high latitudes, then no systematic error due to the Galactic mask is 
expected (\cf \S\ref{sec:testing}). For a few square degrees area, we
notice there are typically no significant gradients in the Galactic RM field
beyond $\vert b \vert \simgreat 30^{\circ}$. This allows us to use the
smooth value safely when analyzing a small area with extragalactic RM
measurements. This value is also consistent with all the abovementioned
values of $D_{\rm RM}$ as obtained by more detailed analyses.

\subsection { The Internal Variation and foreground screens}
\label{subsec:I+f}
We would like to get an estimate for the internal $\RM$ contribution
and for the foreground screen contribution (if the latter exists).
We do not attempt to correct each individual source for the $\RM$
contribution due to the internal and foreground screen terms.
We {\it do not} take the measured $\RM$ and subtract an estimate for 
$\RM_{I+f}$, we rather attempt to estimate the distribution of
$\RM_{I+f}$, and treat it as another noise term.

The difference between the two separate terms in Eq. (\ref{eq:lin_sum})
($\RM_I$ and $\RM_f$ ) emerges from the fact that while the
internal contribution is obtained from the same medium that emits the
polarized light, the foreground screen only serves as a filter for an
already existing signal.
In previous investigations, attempts were made to separate the
two contributions (Burn 1966; Laing 1984), later on it became clear that 
the signature of the two is rather similar (Tribble 1991). 

For our purposes
we are interested in all previous calculations' ``left overs",
their signal is our noise and vice versa. 
For the noise estimate we claim it
is legitimate to take the internal contribution and the foreground
screen contribution together.
We argue that the two can be dealt with simultaneously.
To this end we shall use the other piece of information provided by the
observations -- the polarization degree.

Tribble (1991) shows the connection between the observed depolarization
(as a function of wavelength)
and the observed RM for an extended source. Using this connection, and
provided  the
polarization measurements are available, a direct estimate for the
internal RM can be carried out.
Recall we do not expect any depolarization due to the cosmological RM
since the cosmological coherence we are after is orders of
magnitude bigger ($\simgreat$ a few $\hmpc$) than the source size.

It is interesting to
note, that even if we had perfect information about the
depolarization due to the source structure and foreground screen contribution
and an exact relation between the depolarization degree and the $\RM$, we
could still not subtract this derived $\RM$ from the observed one. This
is due to the fact that the polarization degree doesn't specify the
$\RM$ direction.

Tribble's statistical connection between the observed depolarization and
the observed RM is valid only if the correlation scale of the RM structure
function for the source or the foreground screen is much shorter than the
telescope beam size.
This condition is probably not valid for
individual damped Lyman--$\alpha$ systems, and galaxies that serve as
foreground screens, and for which long range correlation across the telescope 
beam may exist. 

Full justification for neglecting damped Lyman--$\alpha$ systems and galaxies
exists only if we can avoid lines-of-sight that cross such systems.
Otherwise the order of magnitude of an intervening galaxy contribution
is obtained as follows: For a galaxy seen
face-on, where there is a danger of a well ordered magnetic field, the
disc thickness, to which the magnetic field is presumably connected is of
the order of kpc (Simard-Normandin \& Kronberg 1980). 
Observations show magnetic field
magnitude of $\sim\mu$G, mainly in the disc plane.
In order to equate this to a cosmological magnetic field of the order of nG,
and coherent scale of $50 \hmpc$, the  average baryon number density across 
the galaxy should be at
least fifty times higher than the cosmological $\bar n_b$.

However, we do not expect any correlation between the magnetic field orientation
of galaxies along the line-of-sight, or galaxies adjacent (in angle) to each
other. The worst contribution to the RM, could only come from
a random walk of $N_f$ steps where $N_f$ is the number of foreground
screens (intervening systems) along the line-of-sight.
Since one can model the probability for an intersecting system along the
line-of-sight as a function of the source redshift (Welter, Perry, \&
Kronberg 1984; Lanzetta, Wolfe, \& Turnshek 1995), another noise term
can be added. We did not attempt to model this noise term here, because
we believe that the best strategy would be to avoid highly contributing
(\ie damped Lyman-$\alpha$ and Lyman limit) intervening systems.
Low column density Lyman-$\alpha$ system contribute very little to the
RM signal (see Eq. (1.2) in Welter, Perry, \& Kronberg 1984).
An exception may exist, if two sources (or two
separated parts of the same source) cross the same cluster.
In that case the noise of the two
sources ($\RM_{I+f}$) may be correlated, but then we can go back to the
technique that uses the depolarization, in an attempt to remove the noise
correlation.

The best strategy, as stated earlier, would be to avoid intervening systems
altogether. There are two methods for doing this. The first method is
simply to avoid intervening systems by looking for sources that exhibit
no absorption lines in their spectra (we assume spectrography is carried
out anyway, to allow redshift determination).
The existence of a large fraction of quasar  lines-of-sight which do not 
intersect a dense
intervenor is corroborated by 
M{\o}ller \& Jakobsen (1990) and Lanzetta, Wolfe, \& Turnshek (1995)
analyses.

The other method is to put a depolarization limit on the
observations (\eg the one imposed by Tabara \& Inoue 1979). Such a limit
naturally reduces the noise term of $\epsilon_{I+f}$ and does not
affect the correlation signal. As a matter of fact, a sequence of such
depolarization limits, may be helpful in constituting the estimate for
$\epsilon_{I+f}$. There are yet more methods for minimizing the number of sources
that manifest internal RM as we shall point out when we consider real
observations (\S\ref{sec:discuss}).

\subsection{Bayesian Likelihood Analysis}
\label{subsec:bayes}
In the current stage of research we have very little knowledge about the
power spectrum of the cosmological magnetic field. We can hardly even put
limits on its integral value, \ie the rms value of the field.
There are, however predictions regarding the power
spectrum shape and amplitude. In the lack of any observational
preference towards any of these predictions (apart from the limits on
its rms value), we assume that all models are equally probable.
Conventional estimates of the $\RM$ correlation, estimates of the sort
applied to galaxies are not very useful. It is difficult to get 
correct error estimate without simulations, that in turn must assume
some magnetic field power spectrum, moreover in the traditional
correlation calculation procedure, the errors of different bins in $r$
space are correlated. On top of that, the quantity we are after is the
magnetic field power spectrum, and the inversion from the integrated
$\RM$ back to the magnetic field is a non-trivial one.
All of the above lead us to the use of Bayesian statistics as a tool for
finding the best parameters, and their probability, given a model.
In the Bayesian formalism, the a-posteriori probability density of a
model, $\bm$, given the data, $\bd$, is
\begin{equation}
\label{eq:bayes}
\Pr (\bm \vert \bd ) = {\Pr(\bm) \Pr(\bd|\bm) \over \Pr(\bd)} \,.
\end{equation}
As stated earlier, $\Pr(\bm)$, the model probability density is unknown
and therefore assumed uniform for all models. The data probability, in
the denominator is the same for all models, and
therefore serves as a normalization factor. We are thus left with
equivalence between the operation of maximizing the probability of
the model, given the data [$\Pr (\bm \vert \bd )$] and the operation of
maximizing the probability of the data, given the model [$\Pr(\bd|\bm)$].
This equivalence allows us to write down the likelihood function for
$N_{src}$ sources with measured $\RM$.

We begin by constructing the $N_{src} \times N_{src}$ symmetric matrix
$\Upsilon_{ij}$ of the expectation values for the $\RM$ correlation (Eq.
\ref{eq:ups_2}) 
between these sources (as determined by their position). 
We further assume that the errors in each $\RM$ value are
uncorrelated and Gaussianly distributed. The variance of the overall
error is the sum of the quadratures of the various error sources \ie
\begin{equation}
\label{eq:noise_all}
\epsilon _i ^2 = \epsilon _G ^2 + \epsilon _{I+f,i}^2 + \epsilon _{m,i}^2
\, ,
\end{equation}
and the likelihood function becomes
\begin{equation}
\label{eq:like}
{\cal L}  = [ (2\pi)^{N_{src}} \det(\tilde \Upsilon_{ij})]^{-1/2}
  \exp\left( -{1\over 2}\sum_{i,j}^{N_{src}} {\RM_i \tilde \Upsilon_{ij}^{-1}
\RM_j}\right)\,,
\end{equation}
where $\tilde \Upsilon_{ij} = \Upsilon_{ij} + \delta^K_{ij} \epsilon _i ^2 $.\\
The $\chi^2$ statistic is defined as $\chi^2 = -2 {\ln \cal L}$.
The $\chi^2$ statistics as defined, 
is a $\chi^2$ distribution (of $N_{src}$
degrees of freedom) with respect to the {\it data points}, not the
model parameters. This $\chi^2$ is interpreted as a rough estimate for
the confidence levels in the {\it parameter} space.

\subsection{Estimating the Signal-to-Noise Ratio}
\label{subsec:ston}
Before proceeding to the elaborated description of the various tests we
performed with artificial data, we would like to get a rough estimate
for the expected RM correlation and the necessary number of sources for
a high enough signal-to-noise ratio at various scales.
We temporarily abandon the Bayesian approach, and focus on $\Pr (\bm
\vert \bd )$. Recall that the noise in the cross correlation is not
correlated, but may still dominate the signal if not appropriately
averaged over many pairs.
The signal-to-noise ratio for an individual measurement of a pair at
identical redshift, a separation angle $\gamma$, and in the context of
a specific model, is given by 
\begin{equation}
\label{eq:ston}
\left \langle {\Upsilon(z,\gamma) \over \epsilon^2} \right\rangle = {\langle
\Upsilon(z,\gamma) \rangle \over \langle \epsilon^2 \rangle }\, ,
\end{equation}
where in the last equality we assume no dependence between the various
noise terms and the model (cosmology + power spectrum).
The cosmological parameters we choose to use are as follows:
We restrict ourselves to flat cosmology with $\Omega_m=1$, we choose
$H_0 = 100 \kms\, {\rm Mpc}^{-1}$ (but the scaling with $h$ is straightforward).
We take the ionization factor to be $X_n=1$ which according to the 
Gunn-Peterson effect (Gunn \& Peterson 1965) is a reasonable choice, 
and $\Omega_b=0.024 h^{-2}$ (Tytler, Fan, \& Burles 1996), \ie $\bar n_b =
2.6\times 10^{-7}$ cm$^{-3}$.

\begin{figure}[t!]
\hskip0.5truecm {\epsfxsize=2.7 in \epsfbox{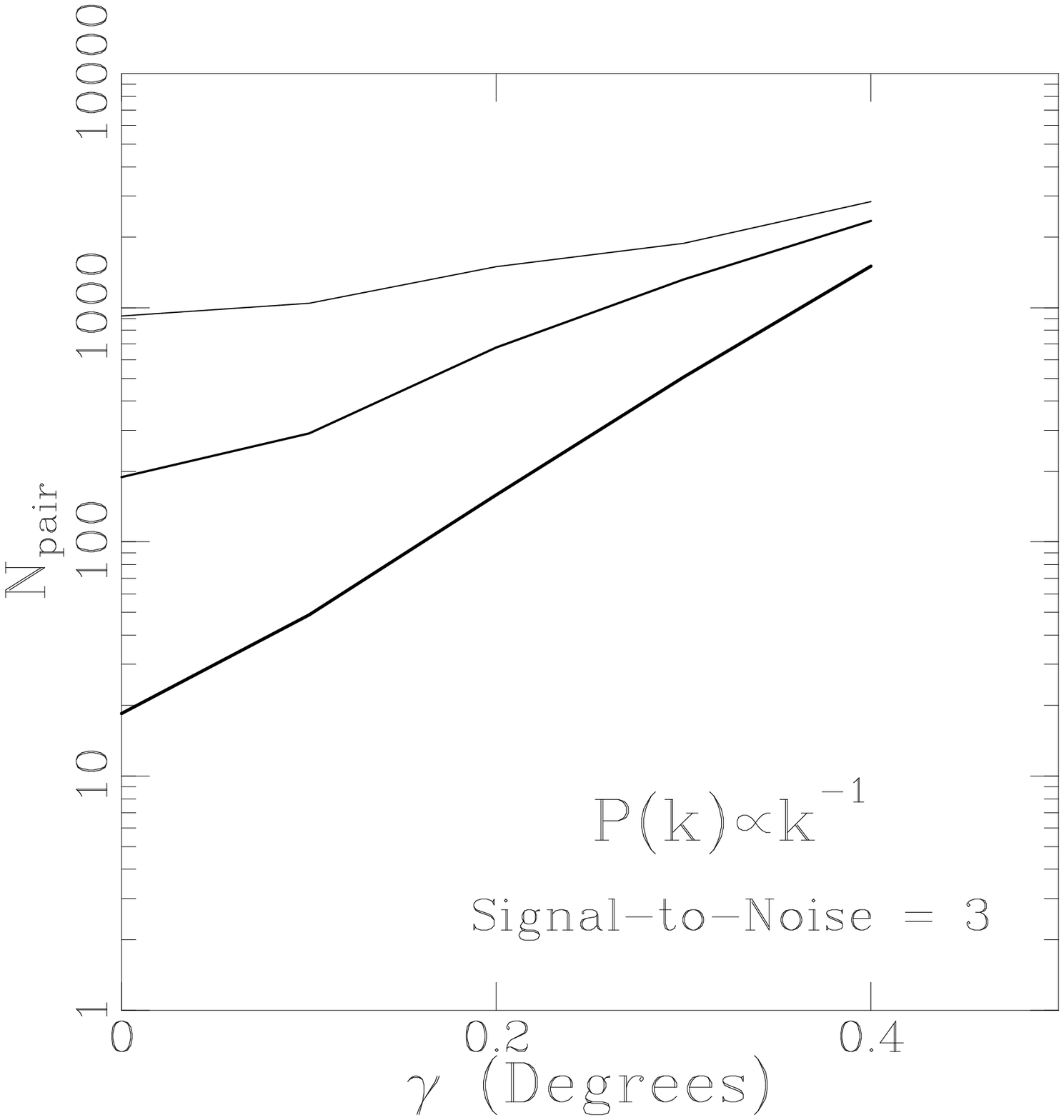}}
\hskip2truecm {\epsfxsize=2.7 in \epsfbox{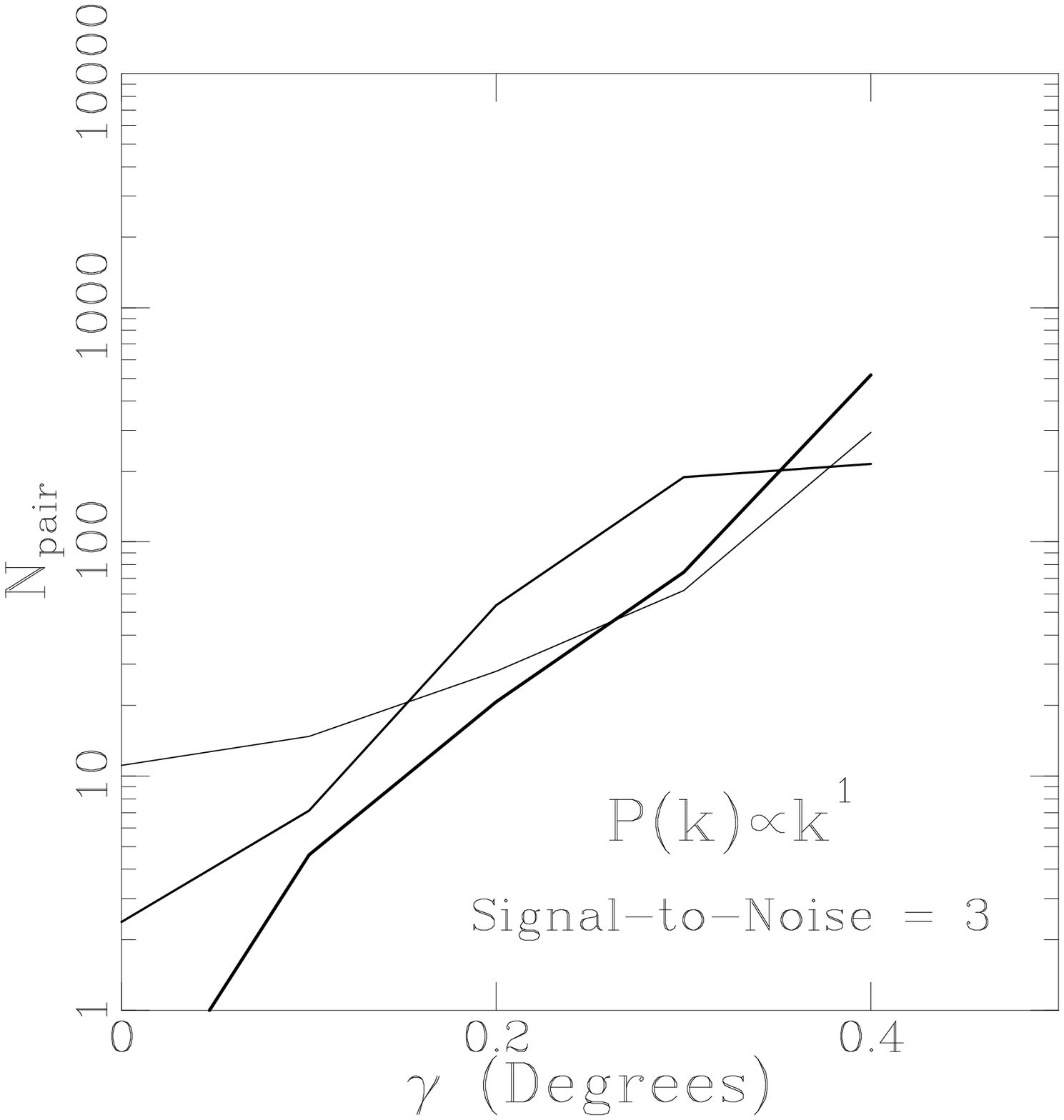}}
\caption{\baselineskip 0.4cm{\trm
Estimate of the number of pairs at separation $\gamma$ needed to
achieve signal to noise
ratio of $3$. The magnetic field is normalized by
$B_{rms}^{TH}(50\hmpc)=1$ nG, and the noise calculation is explained in
the text. Two cases for power index of $-1$ (left) and $1$ (right) are
plotted.
}}
\label{fig:ston}
\end{figure}

For $N_{pair}$ pairs where each source is a member
of one pair only, the signal-to-noise can further be increased by $\sqrt{N_{pair}}$. In figure ~\ref{fig:ston}
we plot
the necessary number of pairs in order to achieve a signal to noise
ratio of $Q=3$. All sources are assumed to be at the same redshift ($2$, $1$,
or $0.5$), we take
the noise values $\epsilon_G=32 \radm2$ (this is the value for 
$\theta_s=1^\circ$). 
For the noise term due to the internal RM and the 
(possible) foreground screen we use the Tabara \& Inoue (1979)
catalog as a guideline. Tabara \& Inoue listed $\sim 1500$ radio sources in
their catalog with $\RM$  $\simless 200 \radm2$. That means a $\sim 3.4
\sigma$ cover, and gives $\sigma_{\RM,obs} \simeq 59 \radm2$. The most
conservative assumption one can make is to attribute all of this $\RM$
to the internal and foreground screen contribution.\\ 
The procedure for deriving the internal and foreground screen contribution
from the depolarization (\cf \S~\ref{subsec:I+f}) bears noise itself and is
correct only statistically. We model its accuracy as a Gaussian with
one quarter the width of the observed RM distribution. Doing so we end up with
$\epsilon_{I+f} = (59^2 + 15^2)^{1/2} \simeq 61 \radm2$. 
This is a conservative choice, and the true value for
$\epsilon_{I+f}$ is probably much smaller.
Note for instance the $\epsilon_{I+f} \simeq 20\, \radm2$ value as obtained by Oren \& Wolfe (1995).
 The measurement errors are neglected
(\cf \S ~\ref{subsec:raw}). 
Two power spectra are considered and they are both normalized to give 
$B^{TH}_{rms}(50 \hmpc) = 1$ nG (the $n=0$ case gives results between the
$n=-1$ and the $n=1$ power spectra). A typical number of pairs for 
achieving a signal to noise ratio of $\sim 3$ in the range $\gamma \simless
0.5$ is between $10$ and $2\times 10^3$ pairs, depending on the power
spectrum. 

This estimate of $Q$ is not accurate. To begin
with, for $N_{src}$ sources we expect less than $(N_{src}^2-N_{src})/2$
statistically independent pairs. We do not expect all sources to reside at the
same redshift, neither do we expect the errors to be identical for all
sources. Furthermore, the separation between the sources is not fixed
, but rather is spread from the arc-minute scale 
to $180^\circ$. 
Due to the need for a realistic error estimate, and a demonstration of an
actual implementation of
the technique we propose, led us to apply it to mock catalogs,
for which the underlying power spectrum is known.

\section{TESTING WITH ARTIFICIAL DATA}
\label{sec:testing}

As a reliable probe for our method we create artificial catalogs of
radio sources for which RM values are measured.
We first generate a Gaussian random field of the magnetic vector potential,
${\vec A}$ 
with the power spectrum $P_A(k) = k^2P_B(k) (=E(k))$. The realization takes
place on a $256^3$ grid points equally spaced in comoving coordinates.
The $P_B(k)$ power spectrum is Gaussianly smoothed on the one grid cell scale.
The amplitude of the field is arbitrary. We then derive the
magnetic field itself by the real space relation $\vec B={\vec \nabla } \times
{\vec A}$ translated to $k$ space \ie ${\vec B}({\vec k}) = -i{\vec k}
\times {\vec A} $. We use the $k$ space symmetries due to the real
nature of the ${\vec A}$ field (no imaginary part), 
and go back to real space to obtain
a divergence-free magnetic field with the desired correlation function,
and periodic boundary conditions.
We check the resultant field by calculating the field
divergence about each and every grid point in real space and get
${\vec \nabla} \cdot {\vec B} / B_{rms} < 5\times
10^{-5}$.

At this point we select the cosmology into which we embed the simulation. 
Our standard choice is identical to the one described in the
signal-to-noise estimate (\S ~\ref{subsec:ston}).
Due to the limited dynamical range, we choose the grid scale to
represent $1.5\hmpc$, and we are therefore confined to a largest
simulation wavelength of $768 \hmpc$.

We then select the source mask, \ie the source distribution across the
sky, and their redshifts.
The two-dimensional distribution is taken to be a Poisson-like
distribution, this choice allows us to have close pairs (unlike
grid selection), but to be as conservative as possible in terms
of the angular correlation function of the sources. Introducing any
correlation as observed for radio sources (Kooiman, Burne \& Klypin
1995; Sicotte 1995; Cress \etal 1995; Loan, Wall, \& Lahav 1997)
can only increase the number of close pairs on small angles, and thus
improve the $\RM$ correlation resultant signal (see fig. \ref{fig:ston}). We do not take explicitly into
account very small angle (sub arc-minute) separations that may exist in
real data sets for resolved extended sources, whose different parts 
can be used as
more than one background source.
We employ two distribution schemes, an all-sky coverage, and a cover of
$150$ deg.$^2$ area centered about the north Galactic pole.

The source redshift distribution is selected according to the $N(z)$ of
radio sources with flux $>35$ mJy (at 4.85 GHz) as calculated by Loan, Wall, \& Lahav (1997)
who used the mean of the theoretical luminosity function models of
Dunlop \& Peacock (1990) and kindly provided their fit.
Loan \etal (1997) $N(z)$ is peaked around $z=1$, and has a 
shape resembling a low redshift truncated Gaussian of width ($2\, \times\, \sigma$)
$\Delta z \simeq 1.9$.

In principle, setting flux limit and sky coverage, sets the
source number. In practice only a sub-set of all radio
sources emit polarized light. A random sample taken from the NVSS survey
(Condon \etal 1994) shows registered polarization angle for $\sim40\%$
of the sources. Refraining from sources with intervening 
systems and large internal RM
may further decrease the number. Since the source number
density in the radio catalogs of this flux limit is a few per square
degree,
when we prepare an ``all-sky" coverage catalog (see below),
we are always way
below the available number of sources. For smaller angular coverage with 
higher source density, one has to further decrease the flux limit. For 
example in the $2.5$ mJy limit we expect $\sim50$ sources per square degree 
(NVSS, Condon \etal 1994), and in the $1$ mJy limit we expect source 
density of about $100$ per square degree 
(the ``FIRST" survey; Becker, White, \& Helfand 1995). 

The $N(z)$ function of Loan \etal (1997)
is therefore interpreted as a normalized selection function, with a
cutoff at the highest redshift of the catalog.
We expect this selection function to provide an underestimate for the number
of sources at high redshift, if a lower flux limit is set. Since sources
at higher redshift typically bear higher signal (without changing the
noise), such an imposed selection is a conservative choice in terms of
the expected signal-to-noise ratio \footnote{The $N(z)$ we choose, which
is peaked in a relatively low redshift, may be useful in mimicking another
effect. As the source redshift increases, the probability for
intervening systems along the line-of-sight increases as well.
If in a sample compilation we try to avoid intervening system, this
attempt translates to a steeper fall-off of the selection function.}.\\

For each selected source, we integrate along the line-of-sight over the
dot product between the magnetic field and the line-of-sight
direction using the periodic boundary conditions. The integration scheme assigns weights to each integration
segment, $k$,  according to
\begin{equation}
\label{eq:weights_b}
w^B_k = { (1+ z_k)^3 \over \sqrt {1-Kr_k^2(z_k)} }\, ,
\end{equation}
in order to account for the cosmological evolution.

We then add the following noise terms to the resultant $\RM$
\begin{enumerate}
\item Internal variation + foreground screens.

We draw this noise term from a Gaussian of $(59 \radm2)^2$
variance (\cf \S~\ref{subsec:ston}). This is the added RM to the
cosmological one. However, we also try to imitate the procedure of
$\epsilon_{I+f}$ evaluation by the depolarization. To this end, for a 
specific value of $\RM^i_{I+f}$ (drawn earlier), we
mimic the recovery of the $\epsilon_{I+f}$ from the depolarization
degree, by listing an $\epsilon_{I+f}$ value scattered about 
$\RM^i_{I+f}$. The scatter has a Gaussian distribution
of $(15 \radm2)^2$ variance. This is the error in the noise estimate.
The ``observer" takes into account the scattered value of the
$\epsilon_{I+f}$ and not the value that was actually added to the
integrated cosmological RM.
\item Galactic mask. 

We use the model-estimate of the Galactic mask, smoothed to $1^\circ$
scale (see \S~\ref{subsec:gal_mask}). Since the smoothed galactic RM differs from the true
$\RM$ to the line-of-sight, we add a $\RM$ drawn from a Gaussian
distribution with the standard deviation of $\epsilon_G=32\radm2$ (Eq. \ref{eq:gal_res}).
\end{enumerate}
At the end of this process we are left with a mock $\RM$ catalog that
consists of source coordinates and redshift, measured $\RM$ value, and the total
error in the $\RM$ value (\ie $\epsilon_{I+f}$).

We proceed by feeding the mock catalog into the maximum likelihood
procedure. 
In the current application, we consider only the right cosmology, \ie
the one in which the simulation was embedded. In principle the
sensitivity to the cosmology choice can be checked as well.
The two variables for the model are the amplitude and the
power index. It is really the combination of the two that the likelihood
procedure constraints best.\\
The statistical significance of the $\chi^2$
distribution (as defined following Eq.
(\ref{eq:like}))
is obtained in the standard way. The number of degrees of
freedom is the source number minus the number of fitting parameters, and the
goodness of fit is calculated by taking into account the value of
$\chi^2_{min}$. If the goodness of fit turns out to be very small
($<0.1$), this is a clear indication for convergence to the wrong
minimum, and vice versa. For the right solution the minimum value for
the $\chi^2$ per degree of freedom (the source number) is typically 
unity within $10^{-3}$ accuracy.

\subsection{Partial Sky Coverage}
\label{subsec:part}
\begin{figure}[t!]
\vskip-1.5truecm
\hskip5truecm {\epsfxsize=2.7 in \epsfbox{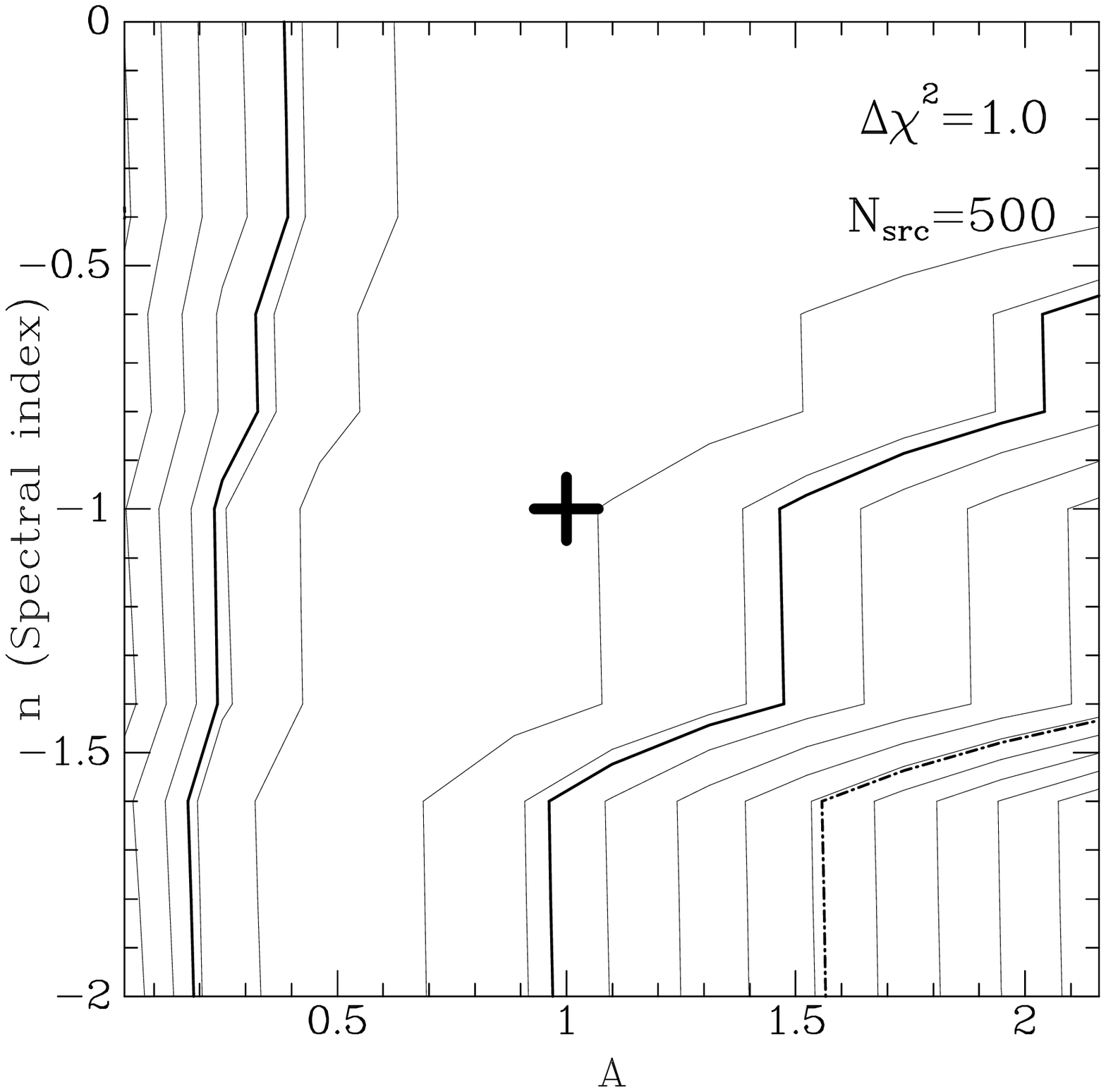}}
\vskip-2truecm
\hskip0.5truecm {\epsfxsize=2.7 in \epsfbox{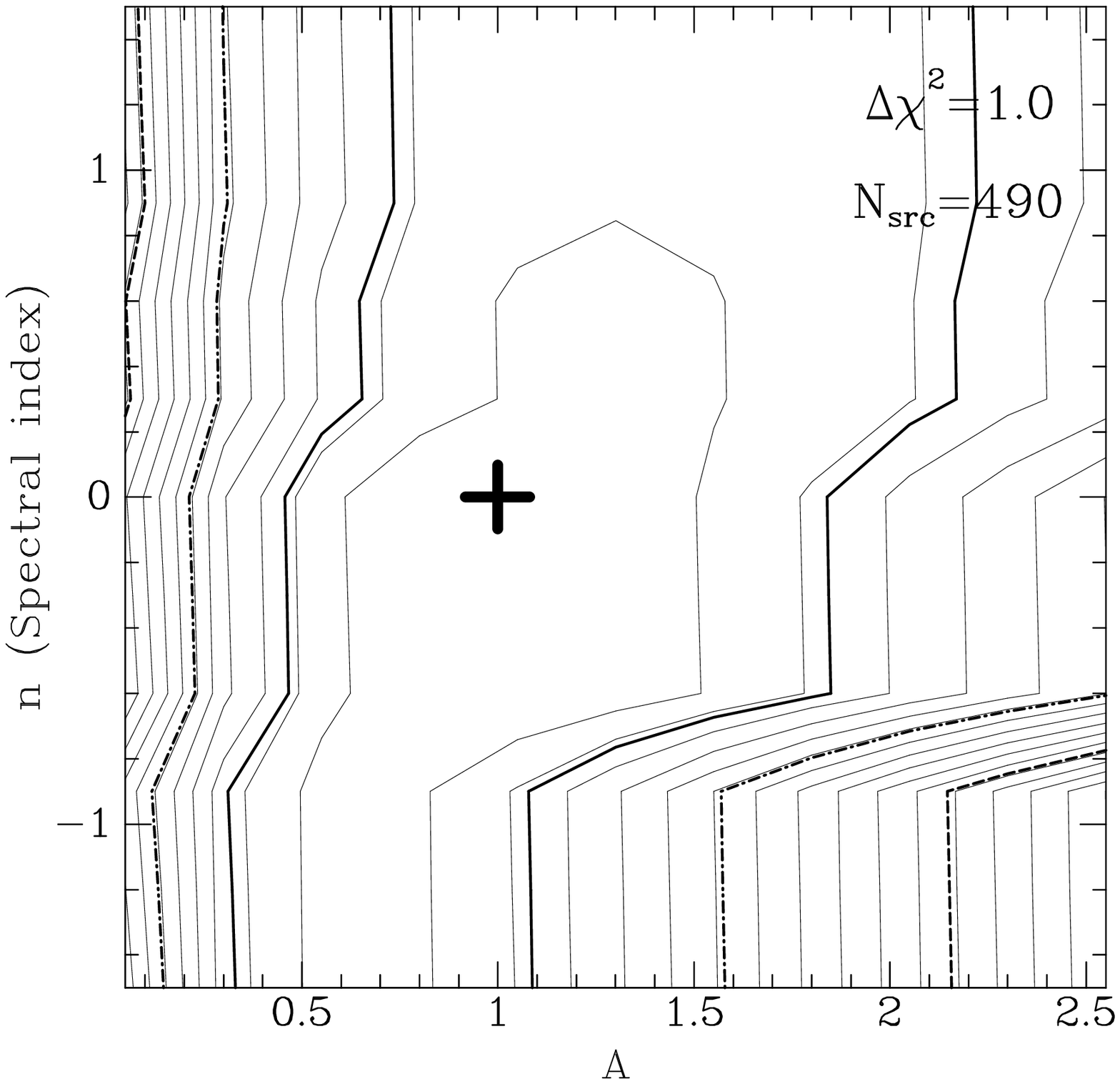}}
\hskip2truecm {\epsfxsize=2.7 in \epsfbox{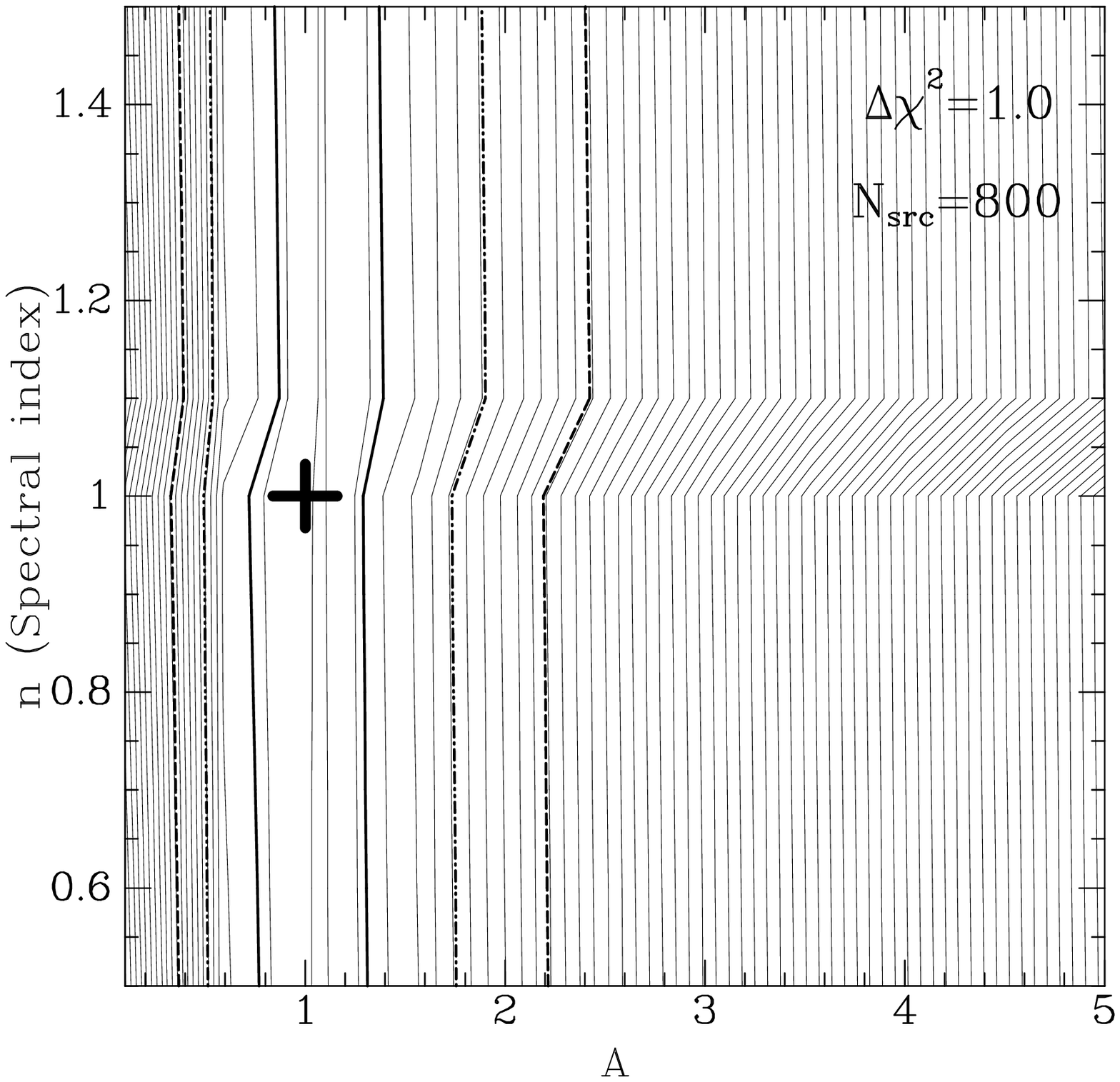}}
\vskip-0truecm
\caption{\baselineskip 0.4cm{\trm
Contour plot of $\chi^2-\chi^2_{min}$ (as defined following Eq.
(\ref{eq:like})) in the amplitude - spectral index plane. The amplitude
is in units of the true simulation amplitude.
The spacing is of $\Delta\chi^2=1$, and confidence
levels for $1,2$, and $3\sigma$ are plotted (bold lines) as if the
contours obey $\chi^2$ statistics in the two-dimensional parameter
space. The true value of the simulation is marked by the cross sign.
All plots are for a part sky coverage of $150$ deg.$^2$ about the
Galactic north pole. The values for $B_{rms}^{TH}(50 \hmpc)$ are $5$,
$2$, and $1$ for $n=-1$, $0$, and $1$ (top, left, right) respectively.
The total number of sources is marked on the plots.
}}
\label{fig:like_part}
\end{figure}

\begin{figure}[t!]
\vskip-1.5truecm
\hskip5truecm {\epsfxsize=2.7 in \epsfbox{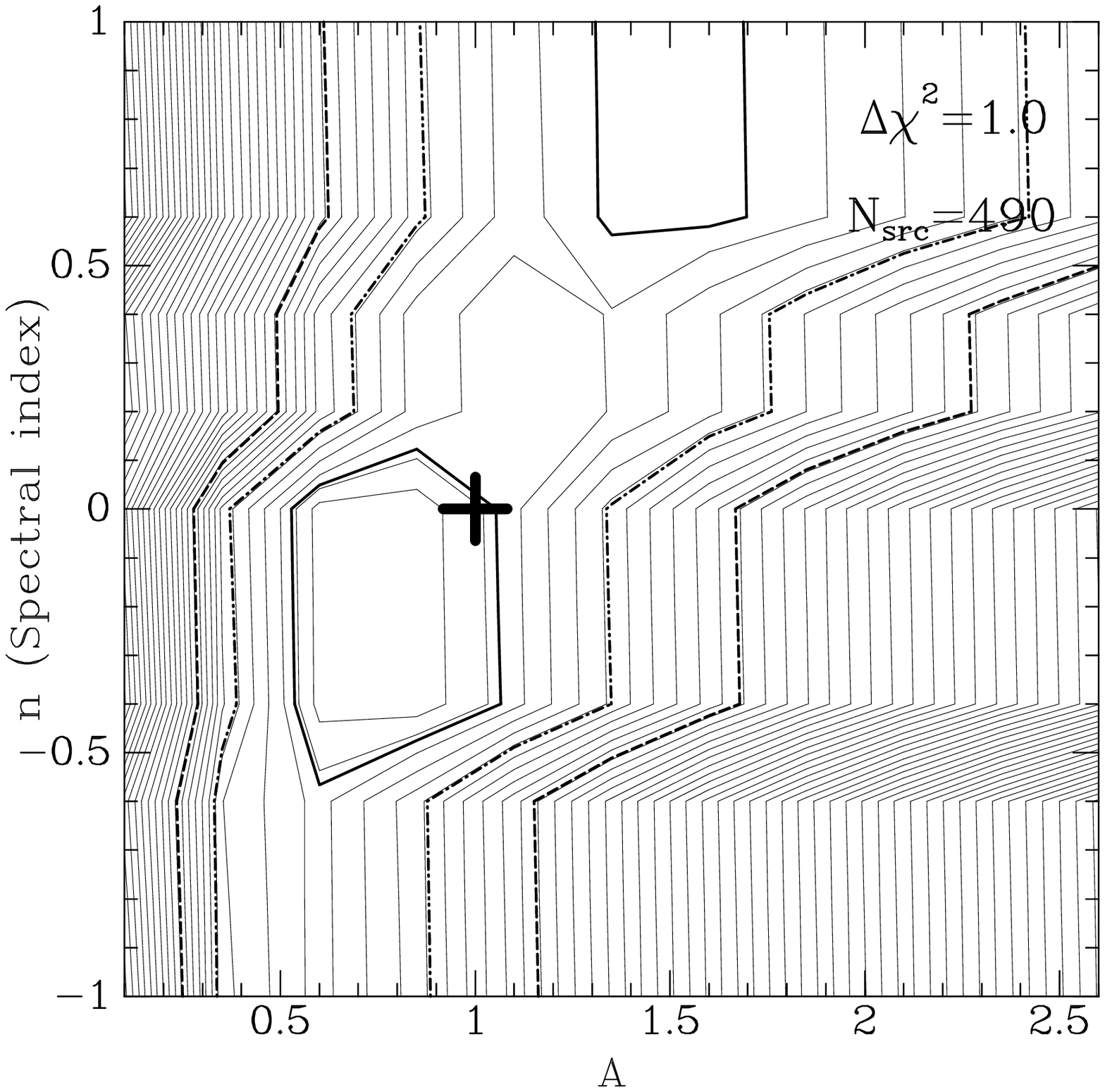}}
\vskip-1.0truecm
\caption{\baselineskip 0.4cm{\trm
Same as the $n=0$ panel in the previous figure (fig.
\ref{fig:like_part}). Here $B_{rms}^{TH}(50 \hmpc)$ is $17$ nG, and
allows a better determination of the parameters.
}}
\label{fig:like_part_high}
\end{figure}

We first explore partial sky coverage mock RM catalogs. 
Figures \ref{fig:like_part} show contour maps of 
$\chi^2-\chi^2_{min}$ with steps
of unity. The thicker lines indicate the approximate $1,2$, and $3\sigma$
levels of the probability function as ascribed by using two parameter $\chi^2$ 
statistics (\cf \S\ref{subsec:bayes}). 
The power spectrum normalization is our standard $B_{rms}^{TH}(50 \hmpc)
= 1-5$ nG, depending on the power spectrum.
We use $500-800$ sources for the part sky and $150$ square degrees coverage.
The different source number and $B_{rms}$ values reflect the fact that
we have fixed the angular sky coverage. That in turn translates differently for
different spectra.  A redshift cutoff of $z<5$ has been applied.\\
Figure \ref{fig:like_part_high} shows a particular example for the case
$n=1$, where we have increased $B_{rms}^{TH}(50 \hmpc)$ to the $17$ nG
level in order to verify the result dependence on $B_{rms}$.

We notice that for all power spectra, the Bayesian approach recovers the
correct value from the artificial catalog (marked by a cross sign) to
within the $1\sigma$ level. 
The method is more robust with respect to the amplitude
determination. The power index is not well bound unless it is very different
from the true one. The method (with the limited number of sources) tends to
recover best the $B_{rms}$ value on a certain scale, which is a combination 
of the power index and the amplitude. Typical values for the ratio
between the true $B_{rms}$ and the minimization result
for $B_{rms}$ are in the range $0.6-1.5$.

Partial sky coverage is preferable in terms of the number of
small separation pairs, and the indifference to the Galactic mask
variation as function of direction.
A collection of several of these small patches is expected to provide
 an even better constrain on the magnetic field power spectrum, but the analysis becomes a bit
more complicated, as one should also include the large scale correlation
between the patches without increasing the noise from the auto
correlation to an intolerable level.

Given the relatively big errors, we desire off-diagonal terms in the
correlation matrix to be as big as possible (\ie small separation).
This is meant to
avoid heavy weight for diagonal terms which are dominated by
noise.
The price for choosing small patches with small sky coverage is limited
ability to constrain the power index.
Under special conditions, this ability can be partially recovered by
turning to an all sky coverage as we discuss in the next section.

\subsection{All Sky Coverage}
\label{subsec:all_sky}
An all sky coverage, with the small correlation scales that are tested
here, is impractical without lowering the error level significantly.
Otherwise the Bayesian analysis will all be dominated by the diagonal
terms with a large noise added to them and one loses the advantage of
the correlation scheme for noise reduction.

All sky coverage is however still feasible if we design the sample very
carefully. In order to minimize the Galactic noise we should observe
sources in proximity to the mask sources direction and avoid the necessity for
smoothing the Galactic mask. Instead, the RM values as deduced from the
mask source population can simply be subtracted from the observed RM values.
The small variation in the Galactic magnetic field on small scales 
(Simonetti, Cordes, \& Spangler 1984; Simonetti \& Cordes 1986; 
Minter \& Spangler 1996) allows this subtraction to take place.

If on top of that one can reduce the noise from internal RM to the level
of $5\, \radm2$ (one can always avoid foreground screens) by multi
wavelength observations, then it becomes sensible to exploit 
an all-sky coverage.

\begin{figure}[t!]
\vskip-1.5truecm
\hskip5truecm {\epsfxsize=2.7 in \epsfbox{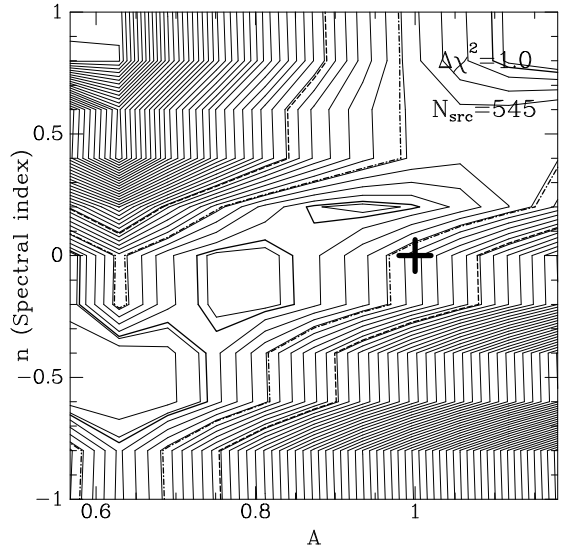}}
\vskip-1truecm
\caption{\baselineskip 0.4cm{\trm
Contour plot of $\chi^2-\chi^2_{min}$
in the amplitude - spectral index plane (same as figure
(\ref{fig:like_part})) for an all-sky coverage of sources in the mask
sources directions. The internal scatter in RM is assumed to be considerably
lower than in the part sky coverage and $B_{rms}^{TH}(50 \hmpc) = 1$ nG.
}}
\label{fig:xlike_all_n=0}
\end{figure}

Figure \ref{fig:xlike_all_n=0} shows an example of a sample taken at the
mask source directions with $\epsilon_{I}=5\, \radm2$ and $B_{rms}^{TH}(50
\hmpc) = 1$ nG. The rest of the parameters are identical to those of the
part sky coverage ($n=1$) of the last section. 
We notice that the power index is somewhat better constrained in an all
sky coverage, especially if the amplitude (or $B_{rms}$) are assumed or
known from some different source.
The true values are recovered to within the $\sim2\sigma$ level.
Moreover such coverage allows us to calculate
all prevailing magnetic fields on the sample scale, expressed as $P_B(k)
= \delta^D(k- {\pi \over R})$ with $\delta^D$ the Dirac delta function
and $R$ the sample depth. Notice, however that using the formalism proposed 
in section \ref{sec:corr},
we cannot take into account power spectra that are not functions of
$\vert k \vert$ alone so this test is a bit weaker than the dipole test.

\section{DISCUSSION AND CONCLUSIONS}
\label{sec:discuss}

We have presented the formalism for the RM correlation function and
demonstrated that by using it we can put limits on
the power spectrum of the cosmological magnetic field.
These limits are more stringent than limits obtained by
any other method that currently exists. Limits of $2-3\sigma$ level for
magnetic field of $\sim1$ nG on inter-cluster scales ($\sim 50 \hmpc$)
can be devised with only $10^2-10^3$ sources. 
The increment of the source number reflects linearly in the $B_{rms}$ limits.

We performed the statistical analysis
following a Bayesian approach that seems to be the most adequate
for this problem.
The correlation method we propose is also less susceptible to systematic
effects such as observational bias (or bias due to evolution) toward 
smaller internal RM measures for higher redshift sources.
Both of these biases tend to lower the expected correlation signal as
the source redshifts increase, whereas a cosmological magnetic field
tends to raise the correlation signal between such sources.

In order to
estimate the statistical significance of limits as derived by the proposed
method, we have simulated a few possible cosmological magnetic fields.
The simulations allow us to have a fair estimate of the errors involved
in the power spectrum derivation.
The analysis depends, however on a few assumptions,
and most crucially on the ability to eliminate or estimate the
foreground screen and internal contribution to the observed RM.

This crucial point can be addressed in the process of the sample
compilation. Reduction of the foreground screen and internal RM
contribution is achievable by applying two cutoff criteria to the
sample.
The first cutoff in minimal polarization degree is meant to avoid 
sources that exhibit large depolarization degree. These sources
are usually interpreted as having a big
non-cosmological contribution to their measured RM.
The second cutoff is for sources that do not follow a $\lambda^2$
dependence of the RM.
Following Laing's (1984) recipe, in order to minimize the number of
sources with internal RM, the wavelength span should allow detection of
deviation from the $\lambda^2$ dependence of the RM for $\phi > \pi/2$.
Should this deviation occur, the source is to be eliminated from the
catalog as a suspect for substantial internal RM contribution.

On top of the abovementioned cutoffs, all sources should preferably be detected for absorption
lines, as a method to eliminate sources with intervening
systems along the line-of-sight to the sources.

The spatial coverage of the sky should be twofold. Small patches of a
few square degrees enable the small separation pair number needed for
noise reduction. Small patches, however, can not reveal large
correlation scales in an effective way. It is thus desirable to compile
a dilute all sky coverage sample, in addition to the high number density
patches off the Galactic plane. The two sampling strategies cover a large range of potential
correlation scales, without the necessity for an all-sky dense sample.
The all sky coverage should preferably look for sources in proximity to
the mask sources directions in order to minimize the Galactic noise term.
Accurate internal RM for these sources should be evaluated through
depolarization measurements in order for
them to be useful for the power spectrum evaluation.
Ultimately, the sample can always be mimicked in terms of the exact
source locations in the framework of all and every detected model.
Exact imitation of the source locations ensures the right correlation
terms in the correlation matrix.

The limits we have derived here are a combination of limits on the free
electron average density, $\bar
n_{e}$, and the magnetic field. If by any other fashion (like HII absorption)
an estimate for $\bar n_e$ can be achieved, then the limits on $B$ will become 
more robust.
The estimate is also a function of the assumed cosmology, and may be
entangled with magnetic field evolution in the post recombination era.
We haven't modeled such evolution in this paper.

This method of RM correlation can further be generalized to the {\it
smoothed} RM correlation. The data can be smoothed on a certain angular
scale, and then either compared to the expectation value predicted by a
model, or inverted numerically to give limits on $B$. The Bayesian
approach, though ceases to be advantageous in the smoothed case, because
even though the noise terms decrease, the smoothing introduces correlation
among the errors, and the statistical interpretation becomes more hazardous.

A nice feature of the proposed analysis is that it is bound to yield
results. These can be either limits on the magnetic field magnitude, or
actual detection of its value. Either way, these results may help in
lifting the curtain over the mystery of the cluster and galactic magnetic 
fields origin.

\acknowledgements
I would like to thank Arthur Kosowsky for extensive discussions and 
helpful assistance, George Blumenthal for valuable insight and
critical reading of the manuscript, and James Bullock for a very carefull
examination of the manuscript.\\
This work was supported in part by the US National Science Foundation
(PHY-9507695).

\end{document}